\documentclass{emulateapj}
\usepackage{graphicx}
\usepackage{epstopdf}
\begin{document}
\title{A Disk Around the Planetary-Mass Companion GSC 06214-00210~\lowercase{b}: \\ Clues About the Formation of Gas Giants on Wide Orbits*}
\author{Brendan P. Bowler,\altaffilmark{1,2} Michael C. Liu,\altaffilmark{1} Adam L. Kraus,\altaffilmark{1,3} Andrew W. Mann,\altaffilmark{1} \\ and Michael J. Ireland\altaffilmark{4,5}}
\email{bpbowler@ifa.hawaii.edu}

\altaffiltext{1}{Institute for Astronomy, University of Hawai`i; 2680 Woodlawn Drive, Honolulu, HI 96822, USA}
\altaffiltext{2}{Visiting Astronomer at the Infrared Telescope Facility, which is operated by the University of Hawaii under Cooperative Agreement no. NNX-08AE38A with the National Aeronautics and Space Administration, Science Mission Directorate, Planetary Astronomy Program.}
\altaffiltext{3}{Hubble Fellow.}
\altaffiltext{4}{Department of Physics and Astronomy, Macquarie University, NSW 2109, Australia}
\altaffiltext{5}{Australian Astronomical Observatory,  PO Box 296, Epping, NSW 1710, Australia}
\altaffiltext{*}{Some of the data presented herein were obtained at the W.M. Keck Observatory, which is operated as a scientific partnership among the California Institute of Technology, the University of California and the National Aeronautics and Space Administration. The Observatory was made possible by the generous financial support of the W.M. Keck Foundation.}

\submitted{ApJ, Accepted (9/22/2011)}
\begin{abstract}

We present Keck/OSIRIS 1.1--1.8~$\mu$m adaptive optics integral field spectroscopy of the planetary-mass 
companion to GSC~06214-00210, 
a member of the $\sim$5~Myr Upper Scorpius OB association. 
We infer a spectral type of L0$\pm$1, and our
spectrum exhibits multiple signs of youth.  The most notable feature is exceptionally strong Pa$\beta$ emission 
($EW$=--11.4$\pm$0.3~\AA) which signals the presence of a circumplanetary accretion disk.  The luminosity 
of GSC~06214-00210~b combined with its age yields a model-dependent 
mass of 14$\pm$2~$M_\mathrm{Jup}$, making it the lowest-mass companion to show
evidence of a disk.  With a projected separation of 320 AU, the formation of GSC~06214-00210~b and other very 
low-mass companions on similarly wide orbits is unclear.  One proposed mechanism is formation at close separations 
followed by planet-planet scattering to much larger orbits.  Since that scenario involves a close encounter  
with another massive body, which is probably destructive to circumplanetary disks, it is
unlikely that GSC~06214-00210~b underwent a scattering event in the past.
This implies that planet-planet scattering is not solely responsible for the population 
of gas giants on wide orbits.
More generally, the identification of disks around young planetary companions on wide orbits offers a novel method 
to constrain the formation pathway of these objects, which is otherwise notoriously difficult to do for individual systems.
We also refine the spectral type of the primary from M1 to K7 and detect a mild (2-$\sigma$) excess 
at 22~$\mu$m using $WISE$ photometry.

\end{abstract}

\section{Introduction}{\label{sec:intro}}

Direct imaging with adaptive optics on large telescopes is beginning to reveal the orbital architecture and 
demographics of extrasolar planetary systems on wide orbits ($>$10~AU).  Several populations of
planetary-mass companions are emerging in this nascent field, including giant planets residing in debris disks 
at moderate separations of $\lesssim$100~AU (HR~8799~bcde: \citealt{Marois:2008p18841}, \citealt{Marois:2010p21591}; 
Fomalhaut~b: \citealt{Kalas:2008p18842}; $\beta$~Pic~b: \citealt{Lagrange:2009p14794}) and 
planetary-mass objects on extremely wide orbits of several hundred AU (e.g., CHXR~73~B: \citealt{Luhman:2006p19659}; 1RXS~J1609--2105~b: \citealt{Lafreniere:2008p14057}; Ross~458~C: \citealt{Goldman:2010p22044}, \citealt{Scholz:2010p20993}).

Despite their growing numbers, the formation mechanisms of these companions remain obscure.  Three plausible 
(and non-mutually exclusive) routes have been proposed: core accretion plus gas capture 
(\citealt{Pollack:1996p19730}; \citealt{Alibert:2005p17987}), 
disk instability (\citealt{Cameron:1978p20284}; \citealt{Boss:1997p18822}), and direct collapse from molecular cloud 
fragmentation (\citealt{Bate:2009p20333}).  \emph{In situ} formation through 
core accretion is unlikely for companions on wide orbits because the timescale to grow massive 
cores at these separations is longer than the observed lifetimes of protoplanetary disks ($\sim$5~Myr; e.g., 
\citealt{Hernandez:2007p387}; \citealt{Evans:2009p22041}; \citealt{Currie:2009p18586}).  On the other hand, 
models of disk instability 
have succeeded in forming gas giants between $\sim$20--100~AU, with specific results depending on the initial conditions and physical
assumptions of the simulations (\citealt{Rafikov:2005p20306}; \citealt{Stamatellos:2008p20312}; \citealt{Boss:2011p22047}).  

There is debate about whether disk instability can account for  planetary-mass companions at separations of 
several hundred AU (\citealt{Boss:2006p18010}; \citealt{Rafikov:2007p20307}; \citealt{Nero:2009p20063}; 
\citealt{Stamatellos:2009p14111}; \citealt{DodsonRobinson:2009p19734}; \citealt{Kratter:2010p20098}; \citealt{Rice:2010p20304}; 
\citealt{Boley:2010p21635}; \citealt{Baruteau:2011p22585}).  Some of the conflicting results 
arise from different approaches to modeling disk heating and cooling as well as
uncertainties in the initial disk masses and surface densities, which are poorly constrained by observations during the
embedded Class 0 and I protostellar phases when this mechanism is most likely to occur.  
Additionally, environmental factors such as 
envelope accretion onto the disk (both steady-state and episodic) are only beginning to be included in simulations (\citealt{Vorobyov:2010p22077}; \citealt{Stamatellos:2011p22430}).  
Despite these difficulties, there is some
evidence that disks around Class 0 objects can be both massive ($>$0.2~M$_{\odot}$) and  
large ($>$200~AU; \citealt{Eisner:2005p22078}; \citealt{Jrgensen:2005p22079}; \citealt{Enoch:2009p22068}; \citealt{Jrgensen:2009p22080}), leaving open the possibility that disk instability can form planets at wide separations.

There have been fewer theoretical studies focusing on the formation of gas giants by direct collapse
from molecular clouds.  
In this scenario, planetary-mass objects form as ejected by-products of a fragmenting  
pre-stellar cloud core.
\citet{Bate:2009p20333} used numerical simulations to follow the gravitational collapse of a 
molecular cloud (see also  \citealt{Bate:2002p20246} and \citealt{Bate:2003p20248}) and found that all stellar and substellar 
objects begin as opacity-limited fragments with masses of a few times that of Jupiter
and subsequently accrete gas, increasing their mass over time.  In these simulations low-mass
brown dwarfs and planetary-mass objects form directly from dense
filamentary cloud structures and from instabilities in disks; in both cases they begin as dynamically unstable 
multiple systems and are ejected from dense regions of gas, halting strong accretion and limiting their masses below
the hydrogen burning minimum mass.  The formation of gas giants at extreme separations may have also proceeded
in this fashion, perhaps by being ejected to wider orbits on faster timescales than brown dwarfs before appreciable accretion 
has occurred.  Alternatively,
as noted by \citet{Bate:2003p20248}, preferential accretion from the primary could also explain the population of both 
brown dwarfs and planetary-mass companions on orbits of several hundred AU.
In this framework of fragmentation plus ejection, planetary-mass objects on wide orbits represent the 
low-mass tail of brown dwarf companion formation.  

In addition to these \emph{in situ} 
formation models, other explanations for giant planets on wide orbits involve formation at close
separations (perhaps through core accretion or disk instability) and subsequent orbital evolution to large separations.
Several possibilities have been proposed, including outward scattering from 
dynamical interactions with another massive planet  
(\citealt{Boss:2006p18010}; \citealt{Debes:2006p21674}; \citealt{Scharf:2009p19732}; \citealt{Veras:2009p19654}) and  
outward resonant migration with another planet while still embedded in a disk (\citealt{Crida:2009p19663}).  
These myriad possibilities make it difficult to identify the formation mechanisms 
of individual systems 
discovered by direct imaging.

The value of this population extends beyond informing formation scenarios; the direct detection of thermal photons 
enables detailed studies of their atmospheres.
In particular, spectroscopy of young low-mass companions 
is providing insight into 
the influence of surface gravity on the atmospheric properties of low-temperature objects.  
\citet{Metchev:2006p10342} find that the young brown dwarf companion HD~203030~B has an earlier spectral 
type than expected from its evolutionary model-derived temperature,
suggesting that gravity may impact the transition 
from L-type dwarfs to T-type dwarfs.  
More recent photometry and spectroscopy of the 
young ($\sim$30~Myr; \citealt{Zuckerman:2011p22621}) HR~8799 planets indicate that they have unusually thick photospheric
clouds, possibly signaling that thick clouds is a general phenomenon of young gas giants 
(\citealt{Marois:2008p18841}; \citealt{Bowler:2010p21344}; \citealt{Currie:2011p21928}; \citealt{Madhusudhan:2011p22575}; \citealt{Barman:2011p22098}).  The emergent spectrum of the $\sim$5~Myr planetary-mass 
companion 2M1207b  shows similar signs of a dusty atmosphere 
(\citealt{Skemer:2011p22100}; \citealt{Barman:2011p22429}).  There is a a growing need to explore this unexpected correlation 
with more objects spanning a range of gravities and temperatures (or equivalently, masses and ages).

Here we present near-infrared spectroscopy of the recently discovered planetary-mass companion GSC~06214-00210~b (\citealt{Kraus:2008p18014}; \citealt{Ireland:2011p21592}), which orbits a member of the Upper Scorpius OB association ($\sim$5~Myr) at a projected separation of 320~AU.  
In $\S$2 we present $J$- and $H$-band spectroscopy of the companion as well as optical and near-infrared spectroscopy
of the primary.  We describe the spectroscopic properties of GSC~06214-00210~b, atmospheric model fits, predictions from evolutionary 
models, and analysis of the primary star in $\S$3.  
In $\S$4 we examine the validity of planet-planet scattering as a migration scenario for GSC~06214-00210~b.
Finally, $\S$5 provides the conclusions of our work.

\section{Observations}

\subsection{Keck/OSIRIS $J$- and $H$-Band Spectroscopy of GSC 06214-00210 b}{\label{sec:osirisspec}}

We observed GSC~06214-00210~b on 9 July 2010 with the OH-Suppressing Infrared Imaging Spectrograph (OSIRIS; \citealt{Larkin:2006p5567}) 
integral field unit with natural guide star adaptive optics on Keck II.  The weather was photometric with good seeing 
(0$\farcs$4-0$\farcs$6 according to the DIMM on CFHT).  We observed GSC~06214-00210~b in the $Jbb$ (1.180--1.416~$\mu$m) 
and $Hbb$ (1.473--1.803~$\mu$m) bandpasses with the 0$\farcs$05 pix$^{-1}$ plate scale, resulting in a lenslet geometry of 16$\times$64, a 
field of view of 0$\farcs$8$\times$3$\farcs$2, and a resolving power ($R \equiv \lambda / \delta \lambda$) of $\sim$3800.
The separation of GSC~06214-00210~b from the primary (2$\farcs$2) is large compared to the measured FWHM of the companion (0$\farcs$11 in $Jbb$; 0$\farcs$13 in $Hbb$) and the seeing disk ($\sim$0$\farcs$5), so 
contamination from the primary is negligible.  We obtained a total of 40 min in $Jbb$ (5~min/exposure $\times$ 4 nodded pairs) and 30 min in $Hbb$ (5~min/exposure $\times$ 3 nodded pairs) with 1$''$ offsets along the long axis of the field of view.  The airmass ranged from 1.35--1.64 during the observations.  
Immediately following our science observations we targeted the nearby A0V star HD~148968 at an airmass of 1.50--1.67.  We also acquired sky frames in both filters before GSC~06214-00210 and after HD~148968.

\begin{figure}
  \resizebox{\textwidth}{!}{\includegraphics{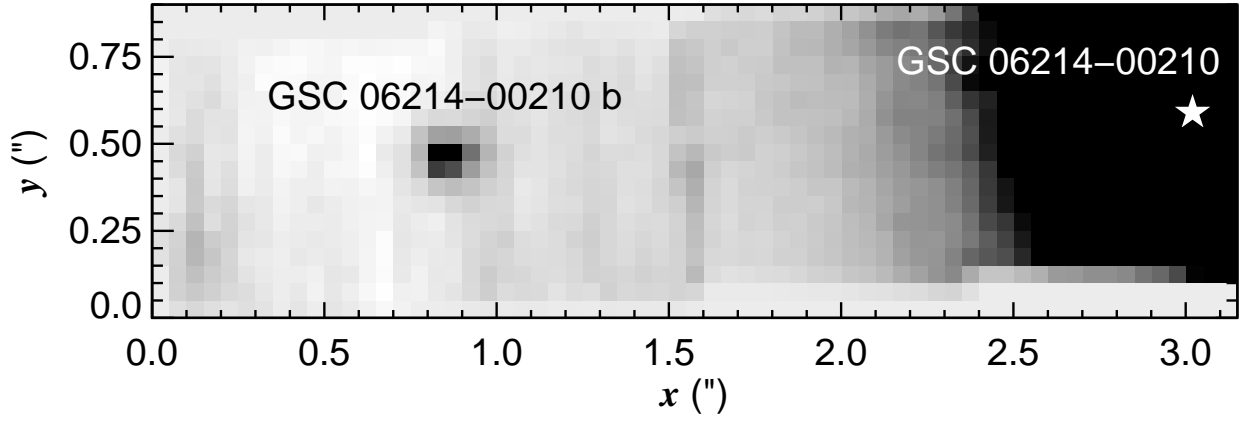}}
  \caption{Median-collapsed $J$-band OSIRIS cube showing GSC~06214-00210 and its companion.  The primary 
  is marked with a white star.  The spaxel scale is 0$\farcs$05 pix$^{-1}$ and the separation between
  the primary and the companion is 2$\farcs$2. \label{osirisdata} } 
\end{figure}

The raw data were flat fielded, sky subtracted, cleaned for cosmic rays and bad pixels, assembled into 3D data cubes using the appropriate rectification matrices, and wavelength calibrated using version 2.3 of the OSIRIS Data Reduction Pipeline.  Figure~\ref{osirisdata} shows an example of a collapsed $J$-band data cube.
The science target and standard star spectra were extracted with aperture radii of 3 and 4 spaxels, respectively.  The individual spectra from 
each band were then scaled to a common level and median combined.  Spectral measurement uncertainties 
were determined by computing the standard errors about the median.  The spectra were telluric corrected with the \texttt{xtellcor\_general} routine in the \emph{Spextool} spectroscopic reduction package (\citealt{Vacca:2003p497}; \citealt{Cushing:2004p501}).  We tested sky subtraction with sky frames (A--sky) and with the nodded science data (A--B) by reducing the data both ways; the influence on the final spectrum was minor.  We chose the 
former (A--sky) for our final spectrum.  

The $J$- and $H$-band spectra were independently flux calibrated using flux ratios for GSC~06214-00210
from \citet{Ireland:2011p21592} together with 2MASS photometry of the primary (\citealt{Skrutskie:2006p589}).
\citet{Ireland:2011p21592} report flux ratios in the MKO system, so we use the
2MASS-MKO conversions from \citet{Leggett:2006p2674} to arrive at  MKO apparent magnitudes for 
GSC~06214-00210~b of $J$=$16.25\pm0.04$ mag and $H$=$15.55\pm0.04$ mag.

\subsection{IRTF/SpeX Near-Infrared Spectroscopy of the Primary GSC~06214-00210}

We observed the primary star GSC~06214-00210 with the SpeX spectrograph (\citealt{Rayner:2003p2588}) in short wavelength cross dispersed mode (SXD) at the 
NASA Infrared Telescope Facility (IRTF) on 29 April 2011 UT.  The seeing reported by the CFHT DIMM was between 0$\farcs$4--0$\farcs$6,
and there were light cirrus clouds during the observations.  We used the 0$\farcs$3 slit aligned to the parallactic angle, yielding a 
resolving power of $\sim$2000,
and obtained a total of 4 min of data (30 s/exposure $\times$ 4 nodded pairs) by nodding along the slit in an ABBA pattern.
After our science observations we targeted the nearby A0V star HD~144925 and obtained calibration frames at a similar airmass.  
The data were reduced with version 3.4 of \emph{Spextool} 
(\citealt{Vacca:2003p497}; \citealt{Cushing:2004p501}).

We also acquired low-resolution ($R$$\sim$100) spectroscopy of the primary with IRTF/SpeX in prism mode on 12 May 2011 UT. 
The weather was poor with complete cirrus coverage so we used the opportunity to verify the spectral slope of our SXD spectrum.  
We used the 0$\farcs$8 slit and obtained 160~s of data (20 s/exposure $\times$ 4 nodded pairs)
at an airmass of 1.38.
Immediately after we observed HD~144925 for telluric calibration.

\subsection{UH~2.2m/SNIFS Optical Spectroscopy of GSC~06214-00210}

We obtained an optical spectrum of GSC~06214-00210 on 16 May 2011 UT under clear
conditions using the SNIFS instrument (\citealt{Lantz:2004p22123}) on the UH~2.2m telescope.  
SNIFS is an optical integral field spectrograph with $R$$\sim$1000--1300 that splits the signal with a dichroic mirror
into blue ($\sim$3000--5200~\AA) and red ($\sim$5200--9500~\AA) channels.  The images are resampled
with microlens arrays, dispersed with grisms, and focused onto blue- and red-sensitive CCDs.

A single 100~s exposure of the science target was sufficient to achieve high S/N ($\sim$200).  
The processing was performed with the SNIFS data reduction pipeline, which is described in detail
in \citet{Aldering:2006p22159} and \citet{Scalzo:2010p22162}.  The processing includes dark, bias, and flat-field 
corrections; assembling the data into 
red and blue 3D data cubes; cleaning them for cosmic rays and bad pixels; 
sky subtraction; extracting the spectra using a
semi-analytic PSF model; and wavelength-calibrating the spectra with arc lamp exposures taken at the 
same telescope pointing as the science data.  Corrections for instrument response, airmass, and telluric lines are based on observations of the Feige~66 standard star using calibrated observations in \citet{Oke:1990p22163}.  
The far ends of the blue and red channels have low QE so to avoid these regions we trim our final spectrum to 
3300--4900 \AA~and 5170--8700~\AA.

\section{Results}

\subsection{Properties of GSC~06214-00210~b}

\subsubsection{Spectral Properties and Classification of GSC~06214-00210~b}

\begin{figure}
  \resizebox{\textwidth}{!}{\includegraphics{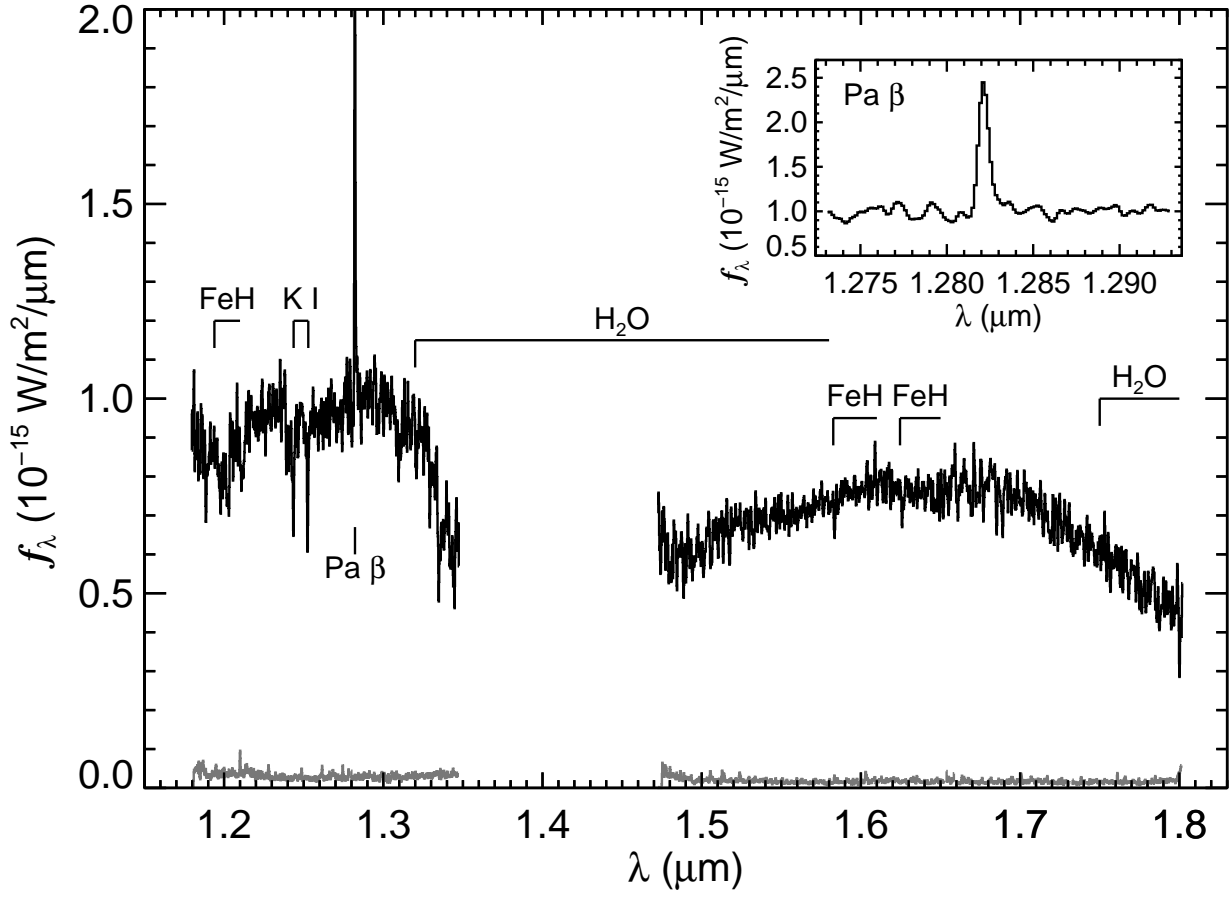}}
  \caption{Flux-calibrated $J$ and $H$ band spectra of GSC~06214-00210~b.  The strong emission line 
  at 1.282~$\mu$m (inset) is Pa$\beta$ ($EW$=--11.4$\pm$0.3~\AA).  The spectrum exhibits
  absorption features typical of late-M/early-L spectral types including FeH, \ion{K}{1}, and H$_2$O. 
   Spectral measurement uncertainties are shown at the bottom in gray. \label{specfig} } 
\end{figure}

Our OSIRIS spectrum of GSC~06214-00210~b is presented in 
Figure~\ref{specfig}.  The most striking feature in the data is the strong Pa$\beta$ emission ($EW$=--11.4$\pm$0.3~\AA)
at 1.282~$\mu$m.  The feature is present is each pair of dithered $J$-band observations 
of GSC~06214-00210~b and spans $\sim$9 spectral elements (Figure~\ref{specfig}, inset), 
so it is not a 
result of a cosmic ray or bad pixel.
One of our $J$-band OSIRIS data sets include the primary (seen in Figure~\ref{osirisdata}; the rest of 
data are dithered to avoid the star) and no Pa$\beta$ emission is observed in that 
spectrum of the star.  Pa$\beta$ emission is 
also absent in
our medium resolution SXD spectrum of the primary obtained $\sim$8.5 months later.  
We therefore
rule out the possibility that the emission observed in the companion is a result of contamination from the primary.

Pa$\beta$ emission in young stars can arise from accretion or outflows
(\citealt{Folha:2001p22164}; \citealt{Whelan:2004p21663}), both of which
imply the presence of a circumplanetary disk around GSC~06214-00210~b.  
The presence of a disk is bolstered by photometry from \citet{Ireland:2011p21592},
who found a red $K$--$L'$ color compared to field M and L dwarfs and 
suggest the excess may originate from thermal disk emission.  Although other parameters such
as metallicity, surface gravity, and dust can also affect the $K$--$L'$ color of late-type objects,
these are unlikely to be the origin of the red color for GSC~06214-00210~b based on the 
expected near solar-metallicity of USco members, a comparison to colors of field objects 
and giants (\citealt{Ireland:2011p21592}), and the minor influence of dust on $K-L'$ colors 
at the M/L transition predicted by models (\citealt{Baraffe:2003p587}; \citealt{Chabrier:2000p161}).  
We note that the $K$--$L'$ color of 1.2 mag is redder than many other young brown dwarfs 
with excesses attributed to disks (e.g., \citealt{Liu:2003p20362}; \citealt{Allers:2009p18424}).
We describe the inferred accretion rate and luminosity in \S \ref{sec:accretion}
and we discuss the implications of a disk in the context of the formation of GSC~06214-00210~b
in \S \ref{sec:formation}.

Other notable features in our spectrum include absorption at $\sim$1.20~$\mu$m from FeH,
1.244 and 1.253~$\mu$m \ion{K}{1} lines, deep H$_2$O steam bands at $\sim$1.4 and $\sim$1.8~$\mu$m,
and FeH bandheads in the $H$ band near 1.6~$\mu$m (Figure~\ref{specfig}).  All of these point to a late-M or early-L spectral type 
(\citealt{Cushing:2005p288}).

Further classification of GSC~06214-00210~b warrants some care since there is no universally adopted
near-infrared classification scheme for young late-type stars and brown dwarfs.
The classification of young brown dwarfs in the literature is inconsistent and often a 
mix of qualitative (comparative) and quantitative (index-based) systems (e.g., \citealt{Reid:1995p22125}; 
\citealt{Kirkpatrick:1999p17664}; \citealt{Martin:1999p19825}; \citealt{Reid:2001p4858}; \citealt{Geballe:2002p19463}; \citealt{Riddick:2007p22157}); when multiple systems are used the adopted
spectral type is usually a (subjectively weighted) average from several schemes.

\begin{figure}
  \resizebox{\textwidth}{!}{\includegraphics{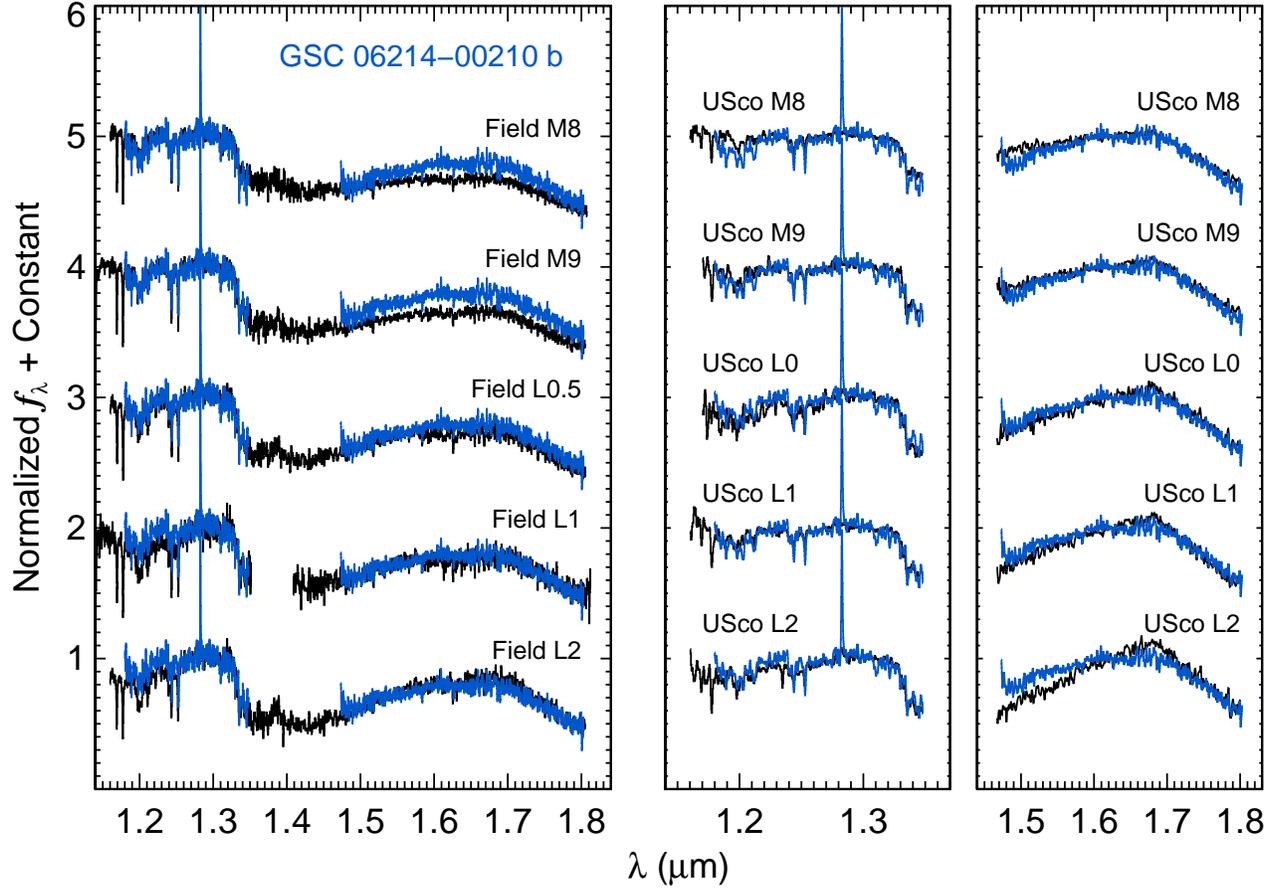}}
  \caption{GSC~06214-00210~b (blue) compared to M8--L2 field objects (left panel) and members of Upper Scorpius (right panels).    
  GSC~06214-00210~b resembles the L0.5 field object, although the depth of the \ion{K}{1} lines are noticeably
  weaker in the younger object.  The $J$ band of GSC~06214-00210~b is well matched to M9--L2 USco objects and the $H$ band to M9--L1 members;
  we adopt a spectral type of L0$\pm$1.  
  Field objects are from the IRTF Spectral Library (\citealt{Rayner:2009p19799}).
   From top to bottom they are Gl~752~B (M8), DENIS-P~J1048--3956 (M9), 2MASS~J0746+2000AB (L0.5), 2MASS~J0208+2542 (L1),
  and Kelu-1~AB (L2).  The USco spectra are from \citet{Lodieu:2008p8698}.
  From top to bottom they are  USco~J155419--213543 (M8), USco~J160847--223547 (M9), USco~J160737--224247 (L0),
  USco~J160723--221102 (L1), and USco~J160603--221930 (L2).  In the left panel the spectrum of GSC~06214-00210~b was 
  smoothed to match the resolving power 
  of the field objects ($R$$\sim$2000).  In the right panels the spectrum of GSC~06214-00210~b and the USco spectra were smoothed to 
  $R$$\sim$1000.  The spectra are normalized to 1.29--1.32~$\mu$m in the left and middle panels and to 1.60--1.65~$\mu$m 
   in the right-most panel.
  \label{fieldusco} } 
\end{figure}

We classify GSC~06214-00210~b primarily based on comparisons to near-infrared 
spectra of field objects and late-type free-floating members of Upper Scorpius (Figure~\ref{fieldusco}).  
The advantage of using field objects is that they are optically typed and the advantage of using USco members
is that they have the same age as GSC~06214-00210~b.  
Spectra of field objects originate from the IRTF Spectral Library (\citealt{Rayner:2009p19799}) and   
those of USco objects from \citet{Lodieu:2008p8698}.  Many of the USco spectra have modest S/N so we first Gaussian smoothed
them from their native resolution of $R$$\sim$1700 to $R$$\sim$1000 and cleaned them for strong single pixel outliers probably
originating form imperfect bad pixel or cosmic ray removal in the original reduction.  For the comparative analysis in 
Figure~\ref{fieldusco} the OSIRIS spectrum of GSC~06214-00210~b was smoothed to the appropriate resolving power for each sample.

The \ion{K}{1} lines in GSC~06214-00210~b are weaker than those in M8--L2 field objects, which is 
a common signature of youth (\citealt{Gorlova:2003p19016}; \citealt{McGovern:2004p21811}; 
\citealt{Kirkpatrick:2006p20500}).  When normalized to 1.29--1.32~$\mu$m, as in 
Figure~\ref{fieldusco}, the best matches to the $J$ band are M8--L0.5 dwarfs.  
The $J$-band in GSC~06214-00210~b is significantly bluer compared to the field L1 and L2 dwarfs.
Together the joint $J$+$H$ spectra appear to best fit the L0.5 dwarf, although the 1.6--1.65~$\mu$m FeH features
are somewhat stronger in the field object.

The best USco matches are M9--L2 objects in the $J$ band and M9--L1 objects in the $H$ band.
Interestingly, the 1.6~$\mu$m FeH features appear to be slightly stronger and the overall $H$ band less triangular in 
GSC~06214-00210~b compared to the USco members.  The depth of the \ion{K}{1} lines are similar for 
GSC~06214-00210~b and the M8--L2 sequence.
Note that the spectral types of some objects from \citet{Lodieu:2008p8698} 
appear to be 1--3 subtypes later than their optical classifications 
(\citealt{Herczeg:2009p18433}; \citealt{Biller:2011p22107}). 

\citet{Allers:2007p66} define a spectral index from 1.50--1.57~$\mu$m based on the depth of the H$_2$O absorption 
that correlates well with optical spectral types 
from M5--L0 and is independent of gravity. This index yields a spectral type of 
M9.5$^{+1.2}_{-1.1}$ for GSC~06214-00210~b, which is consistent with
the comparison to field objects and USco members.  The errors incorporate spectral measurement uncertainties 
and intrinsic scatter in the relation.  Altogether we assign GSC~06214-00210~b a spectral 
type of L0$\pm$1.  Note that this agrees well with the estimate of M8--L4 by \citet{Ireland:2011p21592}
based on colors.

\subsubsection{Effective Temperature and Bolometric Luminosity}{\label{sec:templum}}

\begin{figure}
  \resizebox{\textwidth}{!}{\includegraphics{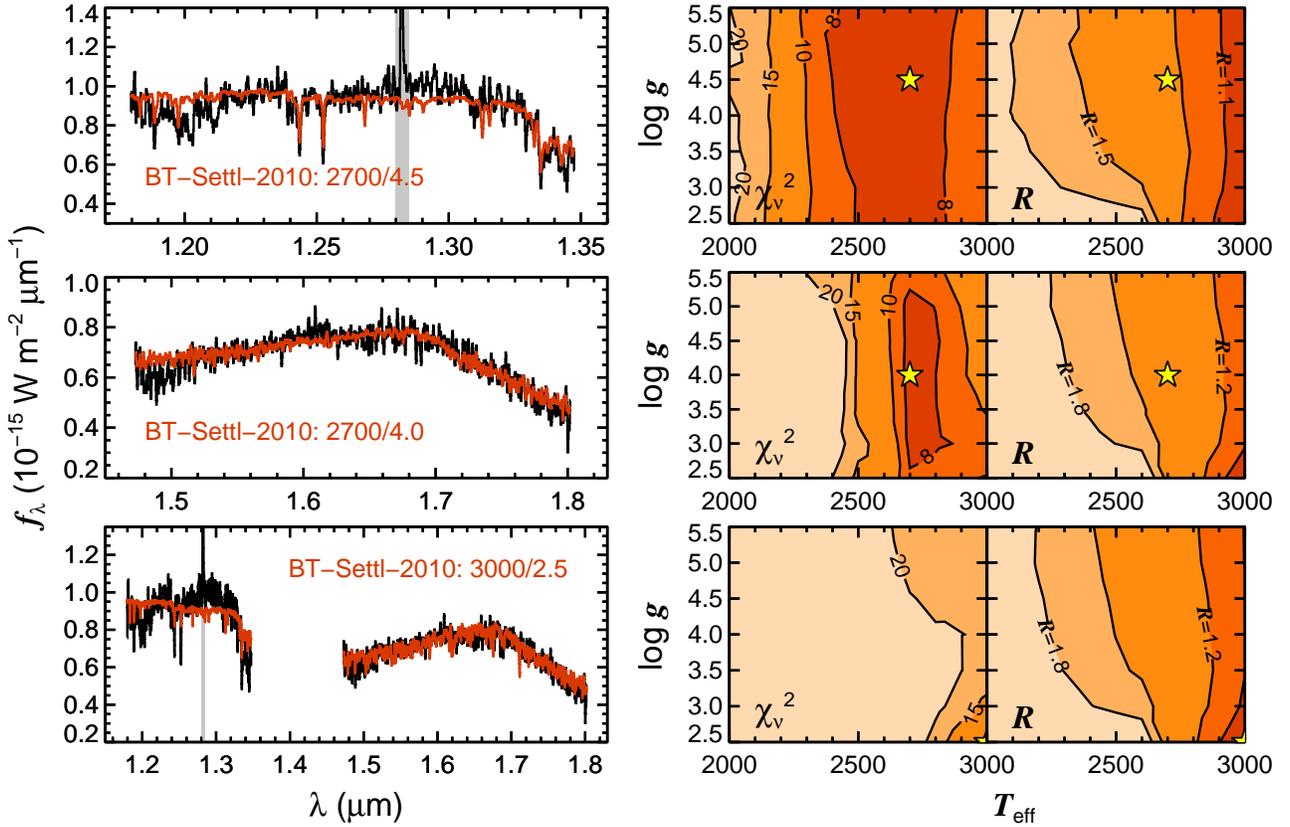}}
  \caption{BT-Settl-2010 atmospheric model fits (red) to our flux calibrated OSIRIS spectrum (black).  $J$ (top), $H$ (middle), and joint $J$+$H$ (bottom) 
  segments are fit separately; the best-fitting models are labeled.  Contour plots map the average reduced $\chi^2$ values and
  inferred radii for each spectral segment based on Monte Carlo simulations of the data, which account for 
  spectral measurement and photometric errors.  We
  exclude the Pa$\beta$ emission line in the fits (gray shaded region).  Surface gravity is poorly constrained compared to effective temperature.  The inferred temperature of 2700--3000~K is significantly higher than evolutionary-model predictions of $\sim$2200~K (Figure~\ref{evtracks}).  The reduced $\chi^2$ contours
  represent values of 8, 10, 15, and 20, and the radius contours represent values of  1.1, 1.2, 1.5, and 1.8 $R_\mathrm{Jup}$.  The best-fit model
  is plotted as a yellow star.  Note that the best-fitting model to the joint $J$+$H$ spectrum is at the edge of the grid.  \label{modfits} } 
\end{figure}

We fit the solar metallicity BT-Settl grid of model atmospheres (2010 version; \citealt{Allard:2010p21477}) to our spectrum of 
GSC~06214-00210~b.  The BT-Settl models incorporate updated water opacity sources, revised solar abundances, and 
a new advanced treatment of dust formation using radiation hydrodynamic simulations.  This new grid reproduces the 
observed colors and SEDs of late-M and L dwarfs better than the Ames-Cond and Ames-Dusty models of \citet{Allard:2001p14776},
which examined the limiting cases of atmospheric dust formation.
The grid spans effective temperatures between 2000--3000~K ($\Delta T_\mathrm{eff}$=100~K) 
and gravities between 2.5-5.5 dex (cgs; $\Delta$log~$g$=0.5).  The fitting procedure relies on $\chi^2$ minimization 
as detailed in \citet{Cushing:2008p2613} and \citet{Bowler:2009p19621}.  To improve the S/N of the data we Gaussian
smoothed our OSIRIS spectrum from its native resolving power of $R$$\sim$3800 to $R$$\sim$2000 and
propagated the measurement errors through the convolution.  We smoothed the models to $R$$\sim$2000 
and fit them to the data in a Monte Carlo fashion 
by randomly generating
synthetic observations based both on the spectral measurement uncertainties and the $J$- and $H$-band photometric uncertainties 
(\S \ref{sec:osirisspec}) used for flux calibration.  
For each trial we determine the best-fitting spectrum and
save the $\chi^2$ values from fitting every model in the grid.  Figure~\ref{modfits} displays $\chi^2$ contour maps averaged over 
100 Monte Carlo trials.  Since the spectra are flux calibrated and the distance ($d$) is constrained to 
145$\pm$14~pc (\citealt{deZeeuw:1999p538}; \citealt{Ireland:2011p21592}),
the radius ($R$) can be calculated from the multiplicative factor ($R^2/d^2$) used to scale the models to the data (see \citealt{Bowler:2009p19621}).
Figure~\ref{modfits} shows the contour plot of the inferred radius for GSC~06214-00210~b, representing averages from the 
Monte Carlo trials and incorporating the uncertainty in the distance estimate.

The $J$ (1.17--1.36~$\mu$m), $H$ (1.45--1.83~$\mu$m), and $J$+$H$ (1.17--1.83~$\mu$m) spectral segments were fit separately.   
To avoid the Pa$\beta$ emission line the 1.280--1.285~$\mu$m region was excluded from the fits. 
The best-fitting model to the $J$ band is $T_\mathrm{eff}$=2700~K/log~$g$=4.5 with $R$=1.22~$R_\mathrm{Jup}$.\footnote{The
standard value for Jupiter's radius is 71,492~km (\citealt{Lindal:1981p20491}), which is the equatorial radius at 1 bar.}
The models match the depth of the \ion{K}{1} lines and 1.35~$\mu$m H$_2$O depth reasonably well, but they fail to 
reproduce the 1.2~$\mu$m FeH feature and the continuum level from 1.27--1.31~$\mu$m.
In the $H$ band the models provide a better match the data.  The best-fit model is $T_\mathrm{eff}$=2700~K/log~$g$=4.0 
with $R$=1.36~$R_\mathrm{Jup}$, which is similar to the result from the $J$ band.  

A fit to the combined $J$+$H$ region 
produces a different result: $T_\mathrm{eff}$=3000~K/log~$g$=2.5 and $R$=1.02~$R_\mathrm{Jup}$.
The quality of the fit is rather poor 
as the $J$ band continuum slope deviates dramatically from the data and the depth of the \ion{K}{1} lines are 
poorly reproduced.  The $H$ band, however, is accurately reproduced by the model.  In this case the
best-fitting model is at the edge of the model grid so it may not represent the global minimum. The 2700~K/4.0~dex model 
that matches the individual bands overestimates the $J$-band flux and underestimates the $H$-band flux
when fit to the entire spectrum.

Another approach to infer effective temperature is to use predictions from evolutionary models based on an object's
age and bolometric luminosity.  The age of the Upper Scorpius association has long been known 
to be $\sim$5~Myr, primarily constrained from fitting isochrones to known members in the HR 
diagram (see \citealt{Preibisch:2008p22220}; \citealt{Slesnick:2008p19814}).  
Inferring the age of a cluster from the HR diagram can be problematic, however, because pre-main sequence 
evolutionary models may have significant systematic errors (\citealt{Hillenbrand:2004p21744}; \citealt{Kraus:2009p19809}) 
perhaps caused in part by  
the influence of episodic accretion (\citealt{Baraffe:2009p21749}; \citealt{Baraffe:2010p21792}).
With these uncertainties in mind, we adopt an age of 5$\pm$2~Myr for GSC~06214-00210~b.

We calculated the bolometric luminosity by integrating an artificial spectrum constructed from the flux calibrated 
OSIRIS  $J$ and $H$ band spectral segments combined with a model spectrum for wavelengths shorter than 
the $J$ band segment (0.001--1.15~$\mu$m), between the $J$ and $H$ segments (1.36--1.45~$\mu$m), and longer 
than the $H$ band segment (1.83--1000~$\mu$m).  At each model-data interface the model was scaled to the data
to create a continuous spectrum.  Uncertainties in the spectra and flux
calibration were accounted for in a Monte Carlo fashion.  To test the sensitivity of the resulting luminosity on the input 
atmospheric model 
temperature we calculated the
luminosity using three BT-Settl models with temperatures of 2000~K, 2500~K, and 3000~K, and a gravity of 4.0 dex.
The influence on the resulting luminosity was negligible ($\sim$0.01~dex).  Based on 1000 Monte Carlo trials
we derived a bolometric luminosity of --3.1$\pm$0.1~dex for GSC~06214-00210~b.

To further verify our luminosity calculation we use the $K$-band bolometric correction calibrated with field objects 
from \citet{Golimowski:2004p15703}.
We arrive at a luminosity of --3.04$\pm$0.10~dex, where the uncertainty takes into account errors in the distance, spectral type,
photometry, and intrinsic scatter in the relation.  In addition we compute the luminosity using parameters predicted from 
evolutionary models (see below; $T_\mathrm{eff}$=2200~K/log~$g$=4.0).  We flux calibrated the BT-Settl, Ames-Dusty, and Ames-Cond
(\citealt{Allard:2001p14776}) models to the mean of the $J$-, $H$-, and $K$-band flux calibration scaling 
factors, accounting for photometric uncertainties in a Monte Carlo fashion.  The resulting luminosities are nearly 
identical at --3.11~$\pm$0.09~dex, only differing by $\sim$0.01~dex among the models.  The radii inferred from the 
scaling factor and the distance (see above) are also virtually the same at 1.89$\pm$0.18~$R_\mathrm{Jup}$.

\begin{figure}
  \vskip -.5 in
  \resizebox{\textwidth}{!}{\includegraphics{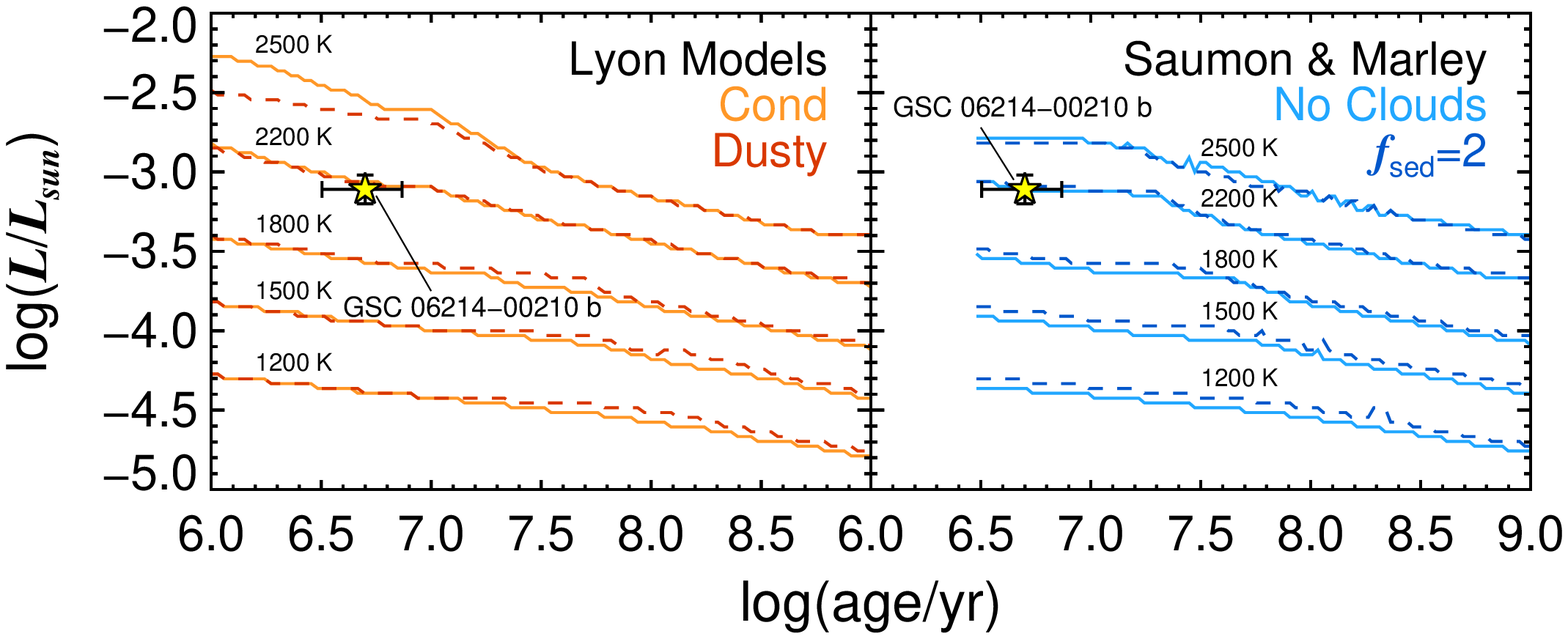}}
  \vskip -2 in
  \resizebox{\textwidth}{!}{\includegraphics{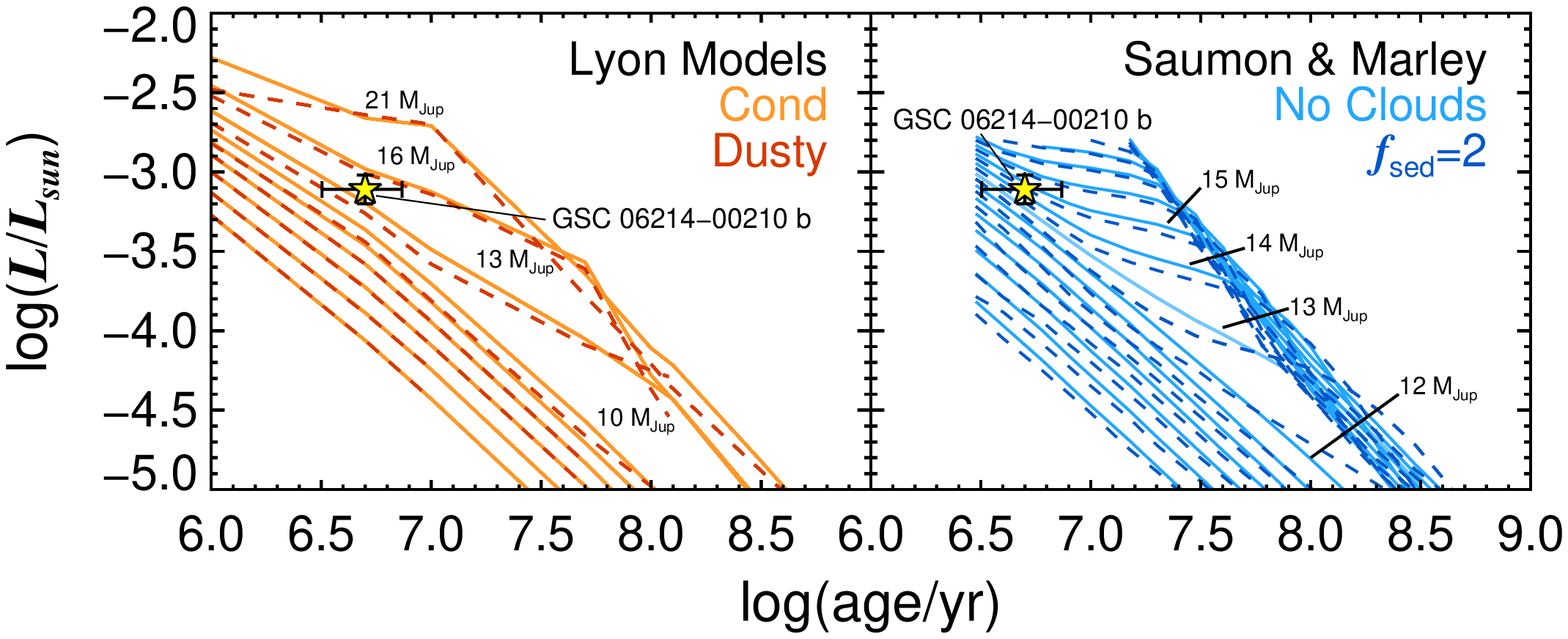}}
  \caption{Evolutionary model-predicted temperature (top) and mass (bottom) of GSC~06214-00210~b.  
  The left panels shows the Lyon Cond and Dusty models of \citet{Baraffe:2003p587} and \citet{Chabrier:2000p161}, while the right
  panels shows the \citet{Saumon:2008p14070} clear and cloudy ($f_\mathrm{sed}$=2) variants.  In the top panels the 
  models are interpolated onto a grid of 
  constant temperatures.  All models yield an effective temperature of $\sim$2200$\pm$100~K for GSC~06214-00210~b 
  with little variation between the cloudy and clear prescriptions or the different models.  
  The Lyon models and the Saumon \& Marley models predict masses within 1 $M_\mathrm{Jup}$ from each other, 
  with the cloudy versions yielding slightly higher masses than the clear variants.  We adopt a mass of
  14$\pm$2~$M_\mathrm{Jup}$ for GSC~06214-00210~b.     \label{evtracks} } 
\end{figure}

Figure~\ref{evtracks} (top panels) displays interpolated temperature tracks as a function of luminosity and age 
for various evolutionary models.  The left panel shows the Lyon Cond and Dusty models 
of \citet{Baraffe:2003p587} and \citet{Chabrier:2000p161}, which demonstrate the 
limiting effects of photospheric dust formation (formation and settling vs. complete retention; 
\citealt{Allard:2001p14776}).
The right panel shows
the evolutionary models of \citet{Saumon:2008p14070} for photospheres with no clouds (``nc'') and 
those with significant amounts of dust (``$f_\mathrm{sed}$=2'').

We determined effective temperatures in a Monte Carlo fashion by interpolating the evolutionary models.
The Lyon models begin at 1~Myr and the Saumon~\&~Marley models begin at 3~Myr, so
we excluded ages younger than those in the analysis.
(The resulting PDFs for the age are truncated Gaussians in linear space).  
The Cond and Dusty models yielded nearly identical results of 2160$\pm$85~K for 10$^4$ Monte Carlo
draws.  The Saumon~\&~Marley models were similar, yielding 2200$\pm$100~K for the 
case with no clouds and 2185$\pm$120~K for the cloudy version.
These values are substantially lower than the those of the best-fitting model atmospheres.

The temperatures inferred from the atmospheric models ($\sim$2700--3000~K) are warmer than previous determinations 
of young M/L transition objects in the literature by $\sim$300--500~K (e.g., \citealt{Bejar:2008p130}; 
\citealt{Kuzuhara:2011p21922}; \citealt{Luhman:2004p22196}).  
A similar result was obtained by \citet{Dupuy:2010p21117}
in their analysis of resolved late-M binaries in the field with dynamical masses.  They found that the most widely used
atmospheric models systematically yield effective temperatures $\sim$250~K higher than the temperatures predicted
by evolutionary models, which are thought to be more reliable because they are less sensitive to missing or 
incomplete opacity sources (\citealt{Chabrier:2000p161}).  Likewise, the radii inferred from the atmospheric model fitting are 
systematically smaller than the value of $\sim$1.95~$R_\mathrm{Jup}$ predicted by evolutionary models 
(\S \ref{sec:accretion}), which is probably a result of overestimated effective temperatures in the fits.

\citet{Luhman:1999p21746} developed an effective temperature-spectral type scale for young M dwarfs 
intermediate between
those of dwarfs and giants.  The calibration was defined to ensure that components of the quadruple system GG~Tau
and members of the young cluster IC~348 were aligned on the same isochrone from the  \citet{Baraffe:1998p160}
evolutionary models in the HR diagram.
\citet{Luhman:2003p22058} revised the scale for M8 and M9 spectral types based on the latest-type members 
of IC~348 and Taurus.  
According to Luhman et al., the effective temperature of a young M9 object is $\sim$2400~K.  
To estimate temperatures beyond the Luhman~et~al. scale, \citet{Allers:2007p66} subtract 
relative offsets taken from the temperature scale of field objects.  This amounts to $\sim$92$\pm$175~K for field
M9--L0 objects based on the relation of \citet{Golimowski:2004p15703} (the uncertainty is from
the rms scatter of the $SpT$-$T_\mathrm{eff}$ relation).  Assuming approximate uncertainties of 100~K for the 
Luhman~et~al. scale, this yields a temperature of $\sim$2310$\pm$200~K, which is consistent with the evolutionary 
model predictions but disagree with those from the atmospheric model fitting.

\subsubsection{Mass}{\label{sec:mass}}

The mass of GSC~06214-00210~b was first estimated to be $\sim$12~$M_\mathrm{Jup}$ by \citet{Kraus:2008p18014} 
from $M_K$ measurements.  It was recently updated to $\sim$12--15~$M_\mathrm{Jup}$ by \citet{Ireland:2011p21592}
based on $JHK$ colors and evolutionary models.  
We refine the mass determination using our luminosity measurement
and a variety of evolutionary models (Figure~\ref{evtracks}, bottom panels).
We use the same procedure to determine the mass from evolutionary models as we use in 
\S \ref{sec:templum} to determine temperature.  
The Cond and Dusty models yield 
13.6$\pm$2.4~$M_\mathrm{Jup}$ and 14.2$\pm$2.4~$M_\mathrm{Jup}$, and the Saumon \& Marley
cloudless and cloudy models yield 14.1$\pm$1.9~$M_\mathrm{Jup}$ and 14.4$\pm$1.8~$M_\mathrm{Jup}$.
These values are in close agreement with previous estimates and hug the border of the brown dwarf/planetary-mass
limit as defined by the deuterium-burning limit ($\sim$13~$M_\mathrm{Jup}$; \citealt{Spiegel:2011p22104}).

\subsubsection{Accretion}{\label{sec:accretion}}

In a study of accretion diagnostics in young brown dwarfs, \citet{Natta:2004p22062} 
found that
Pa$\beta$ luminosity is well correlated with accretion luminosity at low masses.  We use their empirical
relation to derive the accretion luminosity and mass accretion rate for GSC~06214-00210~b.

The equivalent width of the Pa$\beta$ emission line is --11.4$\pm$0.3~\AA, and the flux from that line is 
1.12$\pm$$0.03\times10^{-18}$~W~m$^{-2}$.  Assuming a distance of 145$\pm$14~pc
we find  
log($L_\mathrm{Pa \beta}$/$L_{\odot}$)$=-6.14\pm0.08$ for the Pa$\beta$ line luminosity.  
Applying the empirical relationship between Pa$\beta$ luminosity and accretion luminosity from 
\citet[Equation~2]{Natta:2004p22062} yields an accretion luminosity of 
log($L_\mathrm{acc}$/$L_{\odot}$)$=-4.4\pm1.3$.
The mass accretion rate (\emph{\.{M}}) and the accretion luminosity are related through
\emph{\.{M}}=$L_\mathrm{acc}$$R/GM$, where $R$ is object's radius and $M$ is its mass.
We calculate a radius from evolutionary models in the same fashion as in \S \ref{sec:templum} and \S \ref{sec:mass}
and find $R$=0.20$\pm$0.01~$R_{\odot}$.
This yields a mass accretion rate of log(\emph{\.{M}})=--10.7$\pm$1.3, where \emph{\.{M}} is in $M_{\odot}$ yr$^{-1}$.
The uncertainty is dominated by scatter in the fitted relation from Natta et al. and reduces 
to 0.14~dex when these
are ignored.

\begin{figure}
  \resizebox{\textwidth}{!}{\includegraphics{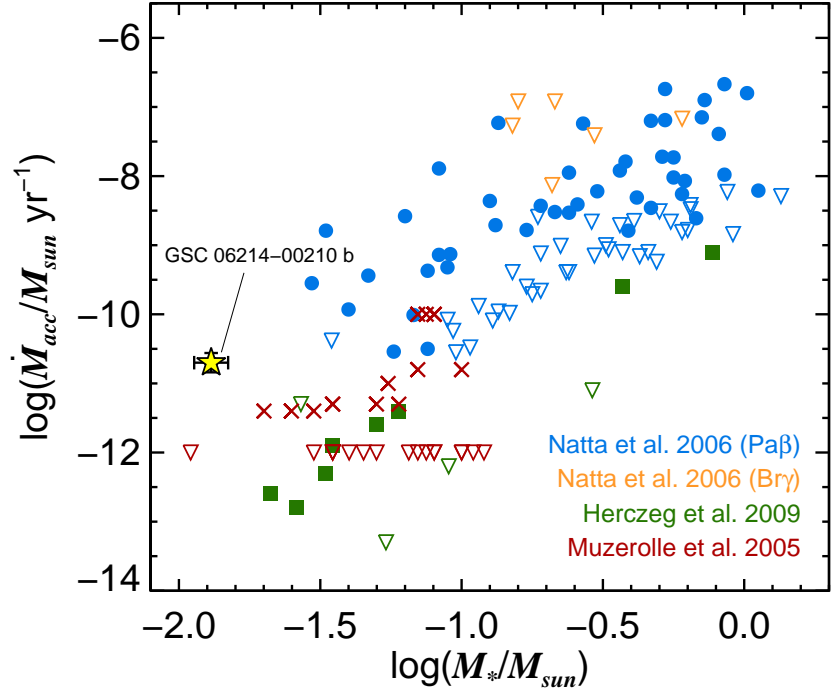}}
  \caption{Mass accretion rates (\emph{\.{M}}) versus stellar mass.   
  The data are from \citet[accretion rates derived using Pa$\beta$ lines are
  shown in blue while those using Br$\gamma$ are in orange]{Natta:2006p22042}, 
  \citet[green]{Herczeg:2009p18433}, and \citet[red]{Muzerolle:2005p55}.  Open inverted triangles
  represent upper limits.  Although the various samples represent different ages and 
  methodologies used to derive mass accretion rates, the data clearly show the strong dependency
  of accretion rate on mass.  The accretion rate of GSC~06214-00210~b appears to be somewhat higher
  than for low-mass brown dwarfs.  The uncertainties shown here in the accretion rate for GSC~06214-00210~b ignore  
  errors in the conversion from $L_\mathrm{Pa \beta}$ to $L_\mathrm{acc}$ from \citet{Natta:2004p22062} since
  we are making a direct comparison to their sample.  The uncertainty increases to $\pm$1.3~dex when the errors are included.
   \label{mdotmstar} } 
\end{figure}

Mass accretion rates are observed to depend strongly on stellar mass and roughly follow 
an \emph{\.{M}}$\propto$$M_* ^2$ empirical relationship 
(e.g., \citealt{Muzerolle:2005p55}; \citealt{Natta:2006p22042}; although see \citealt{Clarke:2006p22428}).
This trend spans two orders of magnitude in mass and six orders of magnitude in mass accretion rate, although
there are over two orders of magnitude of intrinsic scatter in the relation.  
Figure~\ref{mdotmstar} shows the position of GSC~06214-00210~b in the log\emph{\.{M}}-log$M_*$ diagram relative to members
of various star forming regions.  The dependence of accretion rate on mass is clear, and objects
with masses $\lesssim$30~$M_\mathrm{Jup}$ (log($M_*$/$M_{\odot}$)$\lesssim$--1.5) appear to have accretion
rates below $\sim$10$^{-11}$~$M_{\odot}$~yr$^{-1}$.  GSC~06214-00210~b has a somewhat higher accretion rate than
the lowest mass brown dwarfs from \citet{Muzerolle:2005p55} and \citet{Herczeg:2009p18433}, although it is consistent with
the scatter seen at higher masses.

\begin{figure}
  \vskip -1 in
  \resizebox{\textwidth}{!}{\includegraphics{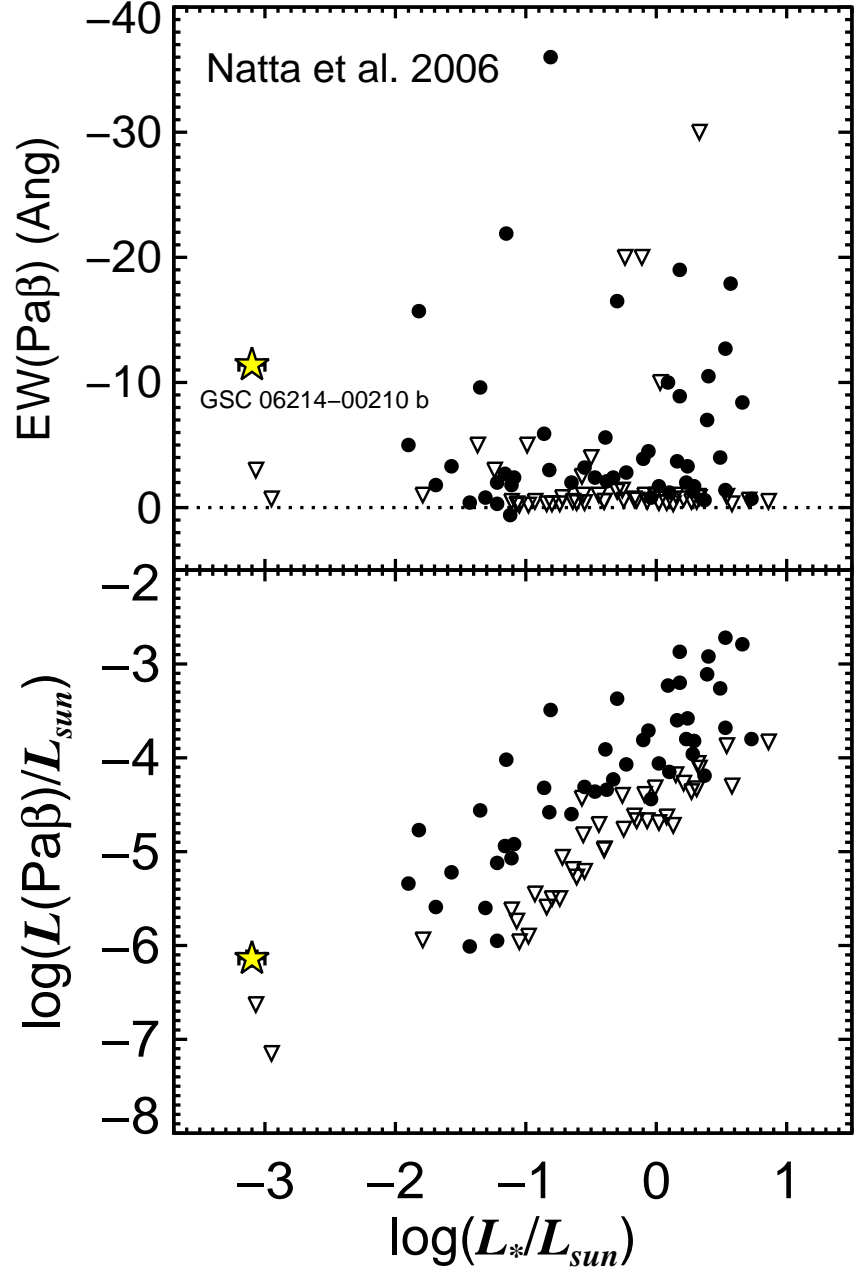}}
  \vskip -1 in
  \caption{Comparison of the Pa$\beta$ $EW$ (top) and line luminosity (bottom) for GSC~02614-00210~b to the \citet{Natta:2006p22042} 
  sample.  Filled circles represent detections and open inverted triangles represent upper limits.  The position of GSC~02614-00210~b
  is marked with a yellow star.
   \label{lstarpabew} } 
\end{figure}

While we have assumed the emission in GSC~06214-00210~b originates from energy released during accretion,
we briefly examine whether it could result from chromospheric activity.  \citet{Short:1998p22581} modeled active M dwarf chromospheres  
and found that weak Pa$\beta$ emission is possible at some chromospheric pressures.  
In Figure \ref{lstarpabew} we show the relative strength of the Pa$\beta$ $EW$ and line luminosity to objects
from the  \citet{Natta:2006p22042} sample of young stars.  The Pa$\beta$ emission strength for GSC~06214-00210~b is 
comparable to the strongest accretors from that sample and is larger than the vast majority of comparison stars.  
We can also examine the correlation between Pa$\beta$ emission and the H$\alpha$ 10\% line width, 
which is a widely-adopted accretion indicator.
Among the sample of young brown dwarfs from \citet{Natta:2004p22062}, four out of the five that show evidence of accretion
based on their H$\alpha$ 10\% widths also exhibit Pa$\beta$ emission.  No objects with Pa$\beta$ emission 
were identified as non-accretors, which together with the unusually strong Pa$\beta$ $EW$ 
supports the notion that the observed emission in GSC~06214-00210~b is a result of accretion.

We note that GSC~06214-00210~b is also likely to have other emission lines.  H$\alpha$ is a good candidate since it correlates 
well with Pa$\beta$ emission in brown dwarfs.  Br$\gamma$ is observed in some, but not all, accreting objects.  For example,
\citet{Natta:2004p22062} found that only two out of eight brown dwarfs with Pa$\beta$ emission showed Br$\gamma$ emission.
However, Br$\gamma$ emission does seem to correlate with the strength of the Pa$\beta$ line, which 
increases the likelihood that GSC~06214-00210~b is emitting at that line. 

\subsection{Properties of the Primary GSC~06214-00210}{\label{sec:primary}}

\begin{figure}
  \resizebox{\textwidth}{!}{\includegraphics{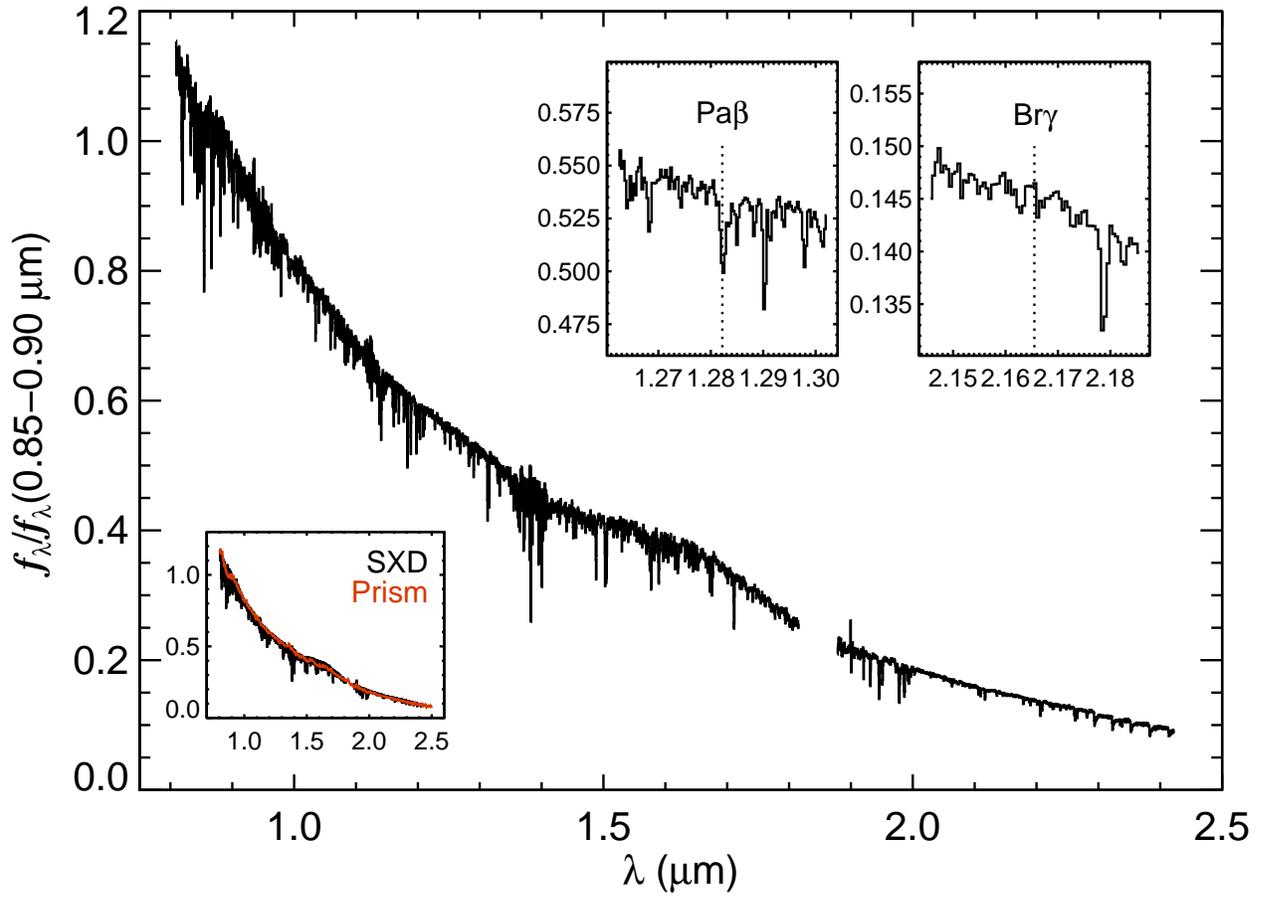}}
  \caption{SXD spectrum of the primary GSC~06214-00210. No Pa$\beta$ or Br$\gamma$ emission lines are 
  evident (upper right insets), which is consistent with mid-infrared photometry showing no indication of a disk.  
  The bottom left inset shows that the SpeX/prism low resolution spectrum (red) agrees well with the SpeX/SXD 
  moderate resolution data.
  \label{primaryfig} } 
\end{figure}

\begin{figure}
  \resizebox{\textwidth}{!}{\includegraphics{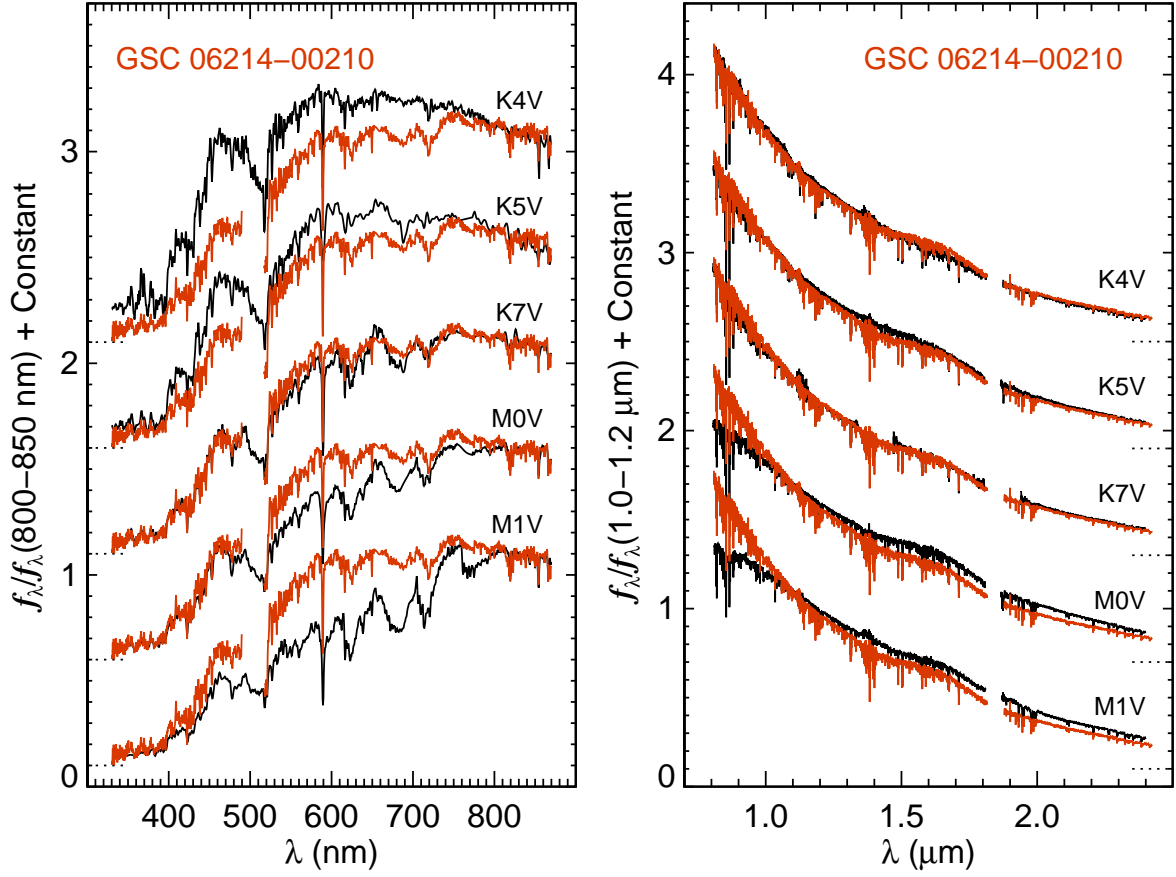}}
  \caption{\emph{Left:} UH~2.2m/SNIFS spectrum of the primary (red) compared to dwarf stars from the
  \citet{Pickles:1998p17673} spectral library. The SNIFS spectrum is a composite of blue (3300--4900 \AA) and red (5170--8700~\AA) channels.  GSC~06214-00210 is well matched by the K7 template.  \emph{Right:} SXD spectrum (red) compared to dwarf stars from the IRTF Spectral Library (\citealt{Rayner:2009p19799}).  From top to bottom the objects 
  are HD~45977 (K4), HD~36003 (K5), HD~201092 (K7), HD~19305 (M0), and HD~42581 (M1).  K4--K7 objects 
  are good matches to our near-infrared spectrum of the primary.  
  We adopt a spectral type of K7$\pm$0.5 for GSC~06214-00210.   \label{primaryoptnir} } 
\end{figure}

The primary star GSC~06214-00210 is a weak-lined T Tauri member of Upper Sco (see \citealt{Ireland:2011p21592} for a review of the literature).
Medium resolution optical spectroscopy by \citet{Preibisch:1998p21909} revealed weak H$\alpha$ emission
($EW$=--1.51~\AA) which can be attributed to chromospheric activity.  
Our 0.8--2.5~$\mu$m SXD spectrum is presented in Figure~\ref{primaryfig};
no emission lines are present.  Compared to late-K and early-M dwarfs from the
IRTF spectral library (\citealt{Rayner:2009p19799}), as shown in Figure~\ref{primaryoptnir} (right panel),
the near-infrared spectrum resembles late-K objects more closely than early-M objects.
We verified that our SXD spectrum was properly reduced by obtaining a low-resolution spectrum with SpeX
in prism mode during a night of poor weather conditions.  The shapes of the spectra are virtually identical
(lower left inset of Figure~\ref{primaryfig}), verifying our SXD spectrum and calling into question the nominal spectral type
of M1 originally assigned by \citet{Preibisch:1998p21909}.  

Overall, the optical spectrum is best matched by that of a K7 dwarf
from the \citet{Pickles:1998p17673} spectral library from $\sim$3300--8700~\AA \ (Figure~\ref{primaryoptnir}, left panel).\footnote{We use the
modification to the MK classification scheme that assigns K7 as an intermediate type between
K5 and M0.}  The shape of the SED and the depth of the TiO absorption bands differ substantially
from those of the M1 spectrum.  We also infer a spectral type using the spectral indices defined by \citet{Reid:1995p22125}, which measure the
depth of various molecular absorption bands in the optical.  
Our value for the TiO5 index (0.916) yields a spectral type
of K5.5.
Since many of the optical absorption bands are gravity-sensitive, 
a slightly earlier spectral type is not unexpected using this index-based scheme.
Similarly, the CaH2 index (0.908) indicates an effective temperature
of $\sim$4200~K using the $SpT$-$T_\mathrm{eff}$ relation from \citet{Woolf:2006p22040}.  All of these diagnostics suggest
a spectral type of late-K rather than early-M, so we revise the spectral type of GSC 06214-00210 from M1 to K7$\pm$0.5.

\begin{figure}
  \resizebox{\textwidth}{!}{\includegraphics{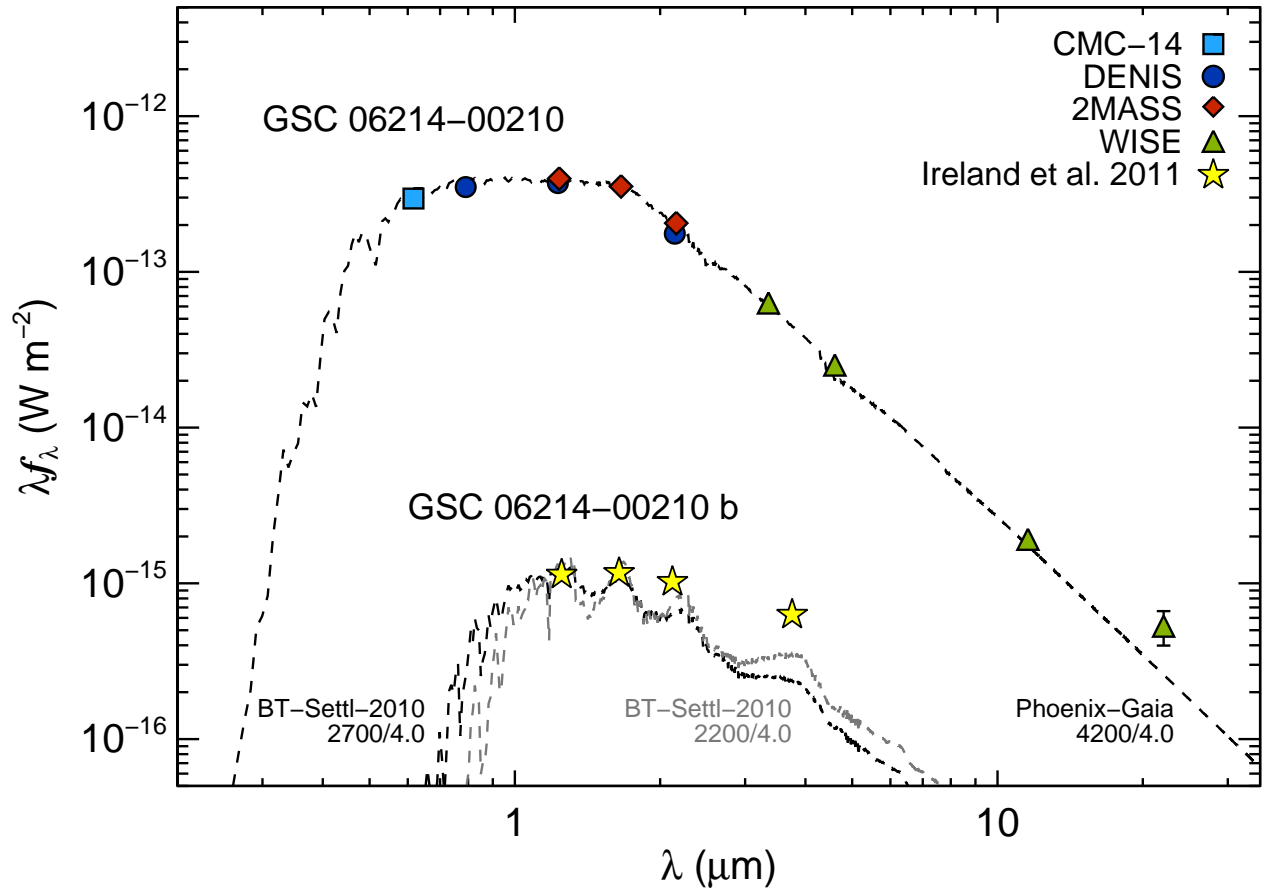}}
  \caption{Spectral energy distributions of  GSC~06214-00210 and its companion.  The $T_\mathrm{eff}$=4200~K/log~$g$=4.0 Phoenix-Gaia model match the 0.6--12~$\mu$m photometry of the primary.  
  For the companion the $T_\mathrm{eff}$=2700~K/log~$g$=4.0 and 
  $T_\mathrm{eff}$=2200~K/log~$g$=4.0 BT-Settl-2010 models are plotted.  The warmer temperature is from fitting 
  atmospheric models and the cooler temperature is the evolutionary model prediction. Both models are flux 
  calibrated to the $J$ band photometry of the companion.  
  The excess flux in $L'$, and perhaps also in $K$ are likely caused by
   thermal emission from a circumplanetary disk.  
  Uncertainties in the photometry are smaller than the symbol sizes except for the $WISE$ 22~$\mu$m point.
  The slight excess at 22~$\mu$m seen in the primary may be caused by a circumstellar disk.  \label{sed} } 
\end{figure}

We determine the effective temperature of GSC 06214-00210 by constructing its SED from 0.6-22~$\mu$m (Figure~\ref{sed}) and comparing it to the 
grid of solar metallicity Phoenix-Gaia model atmospheres (\citealt{Brott:2005p301}).
The photometry originate from the Carlsberg Meridian Catalog 14 (CMC14) for $r'$ band (\citealt{Evans:2002p22242});
the Deep Near-Infrared Southern Sky Survey (DENIS) for $i$, $J$, and $K_S$ bands (\citealt{Epchtein:1997p4581}); 
the Two Micron All Sky Survey (2MASS) for $J$, $H$, and $K_S$ bands (\citealt{Skrutskie:2006p589});
and the Wide-Field Infrared Survey Explorer (WISE) for 3.4, 4.6, 12, and 22~$\mu$m bands (\citealt{Wright:2010p22018}).
Zero point flux densities are from \citet{Stoughton:2002p22243} for CMC14, \citet{Fouque:2000p21781} for DENIS, 
\citet{Rieke:2008p11100} for 2MASS, and \citet{Wright:2010p22018} for WISE.
The models are flux calibrated using the mean scaling factor from the 2MASS $J$, $H$, and $K_S$ bands.
We limit the comparison to models with log~$g$=4.0, which is the approximate surface gravity
from evolutionary models for a K7 star at 5~Myr.
The $r'$, i, and $J$ bands carry the most weight in constraining the effective temperature because the SED
turns over to the Rayleigh-Jeans tail at $\lambda$$\gtrsim$1.6~$\mu$m ($H$ band).
The best match is the 4200~K model, although the 4100~K and 4300~K models
provide decent fits.  Warmer and cooler model temperatures begin to diverge from the photometry, so we assign an 
uncertainty of 150~K for the temperature.  

\begin{figure}
  \resizebox{\textwidth}{!}{\includegraphics{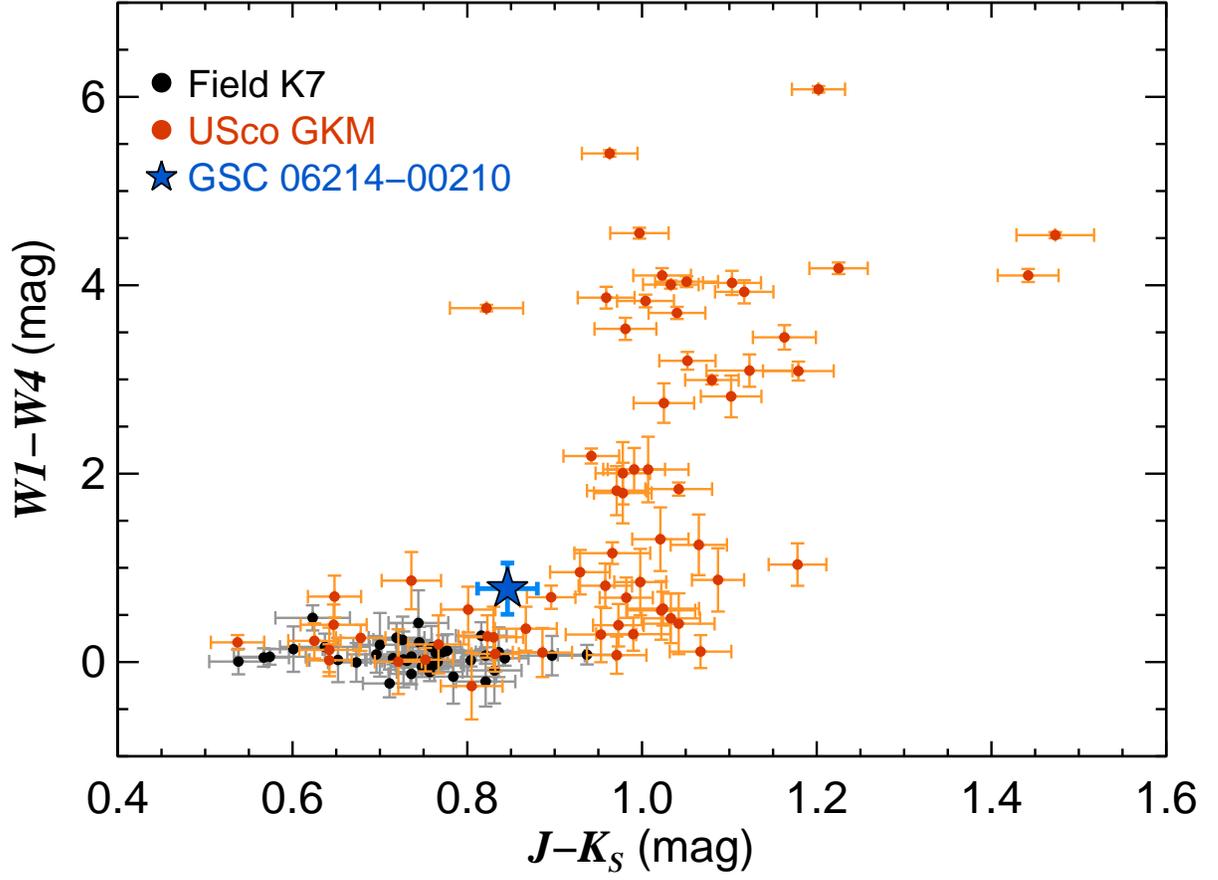}}
  \caption{$W1$--$W4$ vs $J$--$K_S$ diagram for field K7 stars (black filled circles) and 
  Upper Scorpius members with spectral types of G, K, and M (filled red circles).  
  Field stars have $W1$--$W4$ colors near 0.0 mag, while many of the redder USco members
  show an excess at these wavelengths.  
  $WISE$ photometry of GSC~06214-00210 (blue star) suggests
  an excess at the 2-$\sigma$ significance level. \label{wiseccd} } 
\end{figure}

The $WISE$ 22~$\mu$m photometry disagrees with the model in Figure~\ref{sed}, 
suggesting a slight excess for the primary.
In Figure~\ref{wiseccd} we compare the position of GSC~06214-00210 to field K7 dwarfs and other USco members in
the  $W1$--$W4$ vs $J$--$K_S$ diagram.  To isolate a sample of field K7 dwarfs we used a compilation of the revised 
$Hipparcos$ catalog (\citealt{vanLeeuwen:2007p12454}) generated by E. Mamajek (private communication) 
with spectral types and $V$-band magnitudes from
the original catalog (\citealt{Perryman:1997p534}).
An absolute magnitude cut of $M_V$~$>$~5 was imposed to exclude giants.
This yielded 272 K7 dwarfs, which we then fed into the $WISE$ Preliminary Release Source
Catalog query using a search radius of 10$''$. (The PSF FWHM for $WISE$ bandpasses range between 6--12$''$.)  
Among the resulting detections,
we kept only those with the best photometric quality flags in all bands (\emph{ph\_qual}=``A'' or ``B''), 
the best contamination and confusion flags
(\emph{cc\_flags}=``0000''), extended source flags consistent with a point source (\emph{ext\_flg}=``0''), and stable variable flags 
(\emph{var\_flg}$<$5).  This produced 41 objects, which are plotted in Figure~\ref{wiseccd}.
We performed a similar search of the $WISE$ Preliminary
Catalog for known USco members from \citet{Preibisch:2008p22220}.  We queried around 251 USco members with 
GKM spectral types applying the same flags as above; 
67 objects yielded reliable $WISE$ detections.
Field stars and some USco members have $W1$--$W4$
colors of $\sim$0.0$\pm$0.2 mag, although most USco objects show moderate to large excess in that color.  
The $W1$--$W4$ color of 0.78$\pm$0.27 for GSC~06214-00210 indicates a mild (2-$\sigma$) 
excess compared to field K7 stars.

\begin{figure}
  \resizebox{\textwidth}{!}{\includegraphics{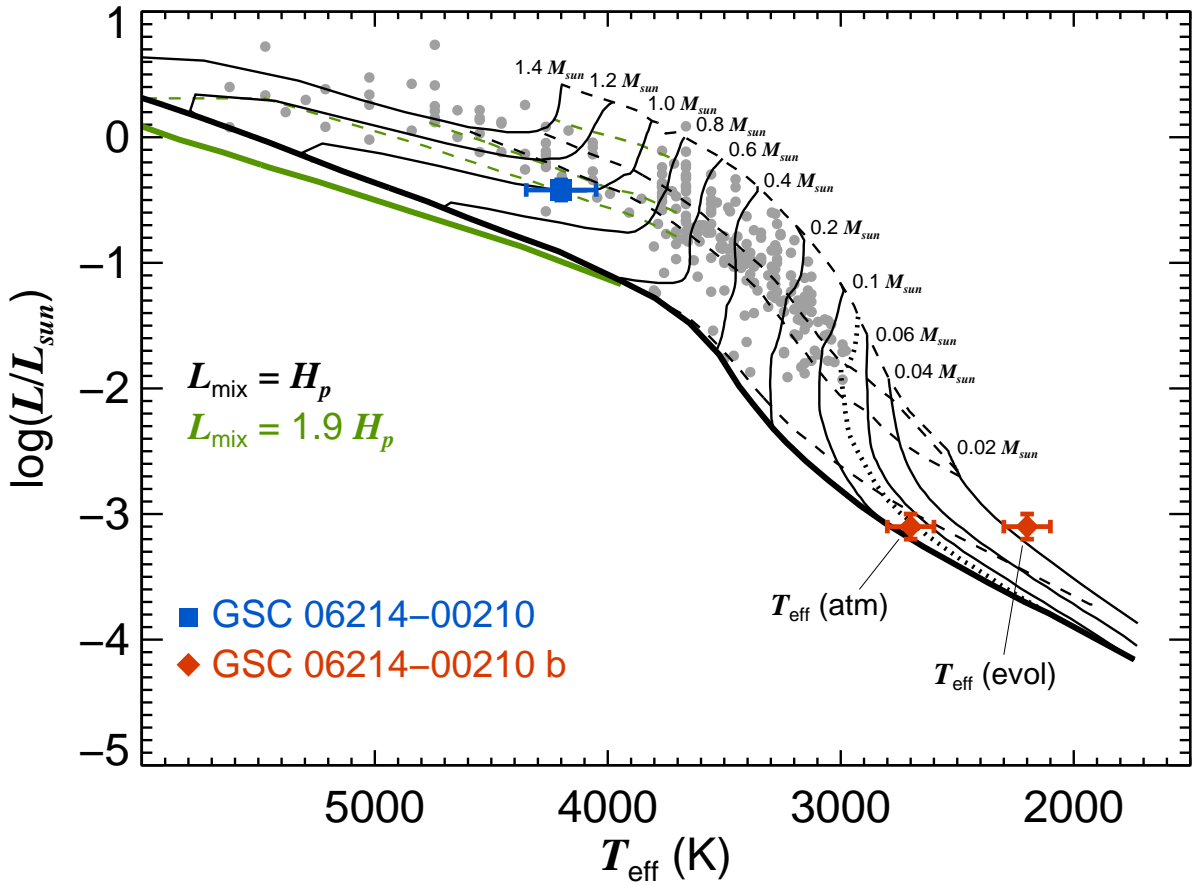}}
  \caption{HR diagram for GSC~06214-00210 and its companion.  The evolutionary models of \citet{Baraffe:1998p160}
  with $L_\mathrm{mix}$=$H_p$ (black) are overplotted with 1, 5, 10, and 100 Myr isochrones (dashed lines) and 
  iso-mass tracks (solid lines) from 
  1.4--0.02~$M_{\odot}$.  The 1~Gyr  main sequence isochrone is shown as a thick line and the hydrogen
  burning minimum mass (HBMM) of $\sim$0.072~$M_{\odot}$ is shown as a dotted line.  The $L_\mathrm{mix}$=1.9$H_p$
  isochrones for 1~Myr, 5~Myr, 10~Myr, and 1~Gyr are shown in green.  The position of 
  the primary indicates a mass of $\sim$1~$M_{\odot}$; our adopted mass of 0.9$\pm$0.1~$M_{\odot}$ is 
  based on an average from several evolutionary models (see text for details).  The position of 
  GSC~06214-00210~b (red) is shown for the warmer atmospheric model-inferred temperature (2700~K) and the cooler
  evolutionary model-derived temperature (2200~K).  The warmer temperature is inconsistent with the young age of the
  system.  Gray circles show members of Upper Scorpius from 
  \citet{Preibisch:2008p22220}.   \label{cmd} } 
\end{figure}

We calculate the luminosity of the primary by integrating the flux calibrated 4200~K/log~$g$=4.0 synthetic spectrum,
yielding $L_*$=0.38$\pm$0.07~$L_{\odot}$.  The error budget is dominated by the uncertainty in the distance
(the uncertainty in effective temperature contributes $\sim$0.01~dex).  
In Figure~\ref{cmd} we show the position of GSC 06214-00210 in the HR diagram.  The pre-main sequence evolutionary
models of \citet{Baraffe:1998p160} for $L_\mathrm{mix}$=$H_p$ are displayed from 0.02--1.4~$M_{\odot}$ with isochrones 
from 1~Myr to 1~Gyr (black).  We also include the $L_\mathrm{mix}$=1.9$H_p$ 1~Myr, 5~Myr, 10~Myr, and 1~Gyr isochrones 
for higher masses ($\geq$0.6~$M_{\odot}$).
Gray circles show the positions of Upper Scorpius low- and intermediate-mass members
in \citet[Tables 1 and 2]{Preibisch:2008p22220}.   
The position of GSC 06214-00210 is consistent with the scatter exhibited by 
other Upper Scorpius members, although it sits at a somewhat older isochrone than the nominal 5~Myr age of the
complex.  For the prescription in Figure~\ref{cmd} with $L_\mathrm{mix}$=$H_p$, the isochronal age is 
$\sim$16~Myr with a 1-$\sigma$ range of 13--25~Myr.  For $L_\mathrm{mix}$=1.9$H_p$, which is required to fit the Sun, 
the inferred age is $\sim$10~Myr with a range of 6--16~Myr.  Note that the isochronal age of low-mass USco members ($\sim$5~Myr)
diverges from the cluster age based on the $L_\mathrm{mix}$=$H_p$ prescription ($\sim$10~Myr), and instead is more
consistent with the age inferred from the $L_\mathrm{mix}$=1.9$H_p$ grid.

We estimate the mass of GSC 06214-00210 from its position on the HR diagram together with predictions from pre-main 
sequence evolutionary models.  
The BCAH98 models imply a mass
of $\sim$1.0~$M_{\odot}$ (Figure~\ref{cmd}).  We also compare it to the evolutionary models of \citet{DAntona:1994p22345},
which yield a mass of $\sim$0.8~$M_{\odot}$, and \citet{Palla:1999p21795}, which yield  $\sim$0.9~$M_{\odot}$.  Given these
moderate systematic differences, we adopt a mass of 0.9$\pm$0.1~$M_{\odot}$.  Note that this
is significantly higher than the value of 0.6$\pm$0.1~$M_{\odot}$ that has been cited in previous work using the later 
M1 spectral type.

\section{Discussion}
\subsection{Did GSC~06214-00210~b Experience a Scattering Event?}{\label{sec:formation}}

As described in \S \ref{sec:intro}, several explanations can account for the observed population of planetary-mass
companions orbiting stars at several hundred AU.  Unfortunately, these models make few unique or testable predictions,
and there is considerable debate about formation scenarios in the literature.
Here we focus on the viability of one explanation, planet-planet scattering, in the context of our results for GSC~06214-00210~b.

The formation of closely packed planetary systems with two or more giant planets naturally leads to dynamical
interactions which can significantly alter the orbits of one or more components.  These close encounters  
produce a wide range of outcomes including (but not limited to) collisions between planets, accretion onto the star, 
rearrangement to new stable or quasi-stable orbits, scattering to close separations, scattering to highly eccentric
wide orbits, and complete ejection (``ionization'').
There is growing observational evidence that scattering is an important phenomenon in
extrasolar planetary systems.  It has been invoked to explain the population of planets with high eccentricities
found in radial velocity surveys (e.g., \citealt{Rasio:1996p19781}; \citealt{Ford:2008p19033}; \citealt{Juric:2008p22245}), 
the distribution of Rossiter-McLaughlin spin-orbit measurements (\citealt{Morton:2011p22267}), 
and, more recently, the possible abundance of unbound gas giants inferred from 
microlensing surveys (\citealt{Sumi:2011p22269}).

We examine a hypothetical scattering event for GSC~06214-00210~b by assuming it was formed much closer than its present 
location through
conventional means (core accretion or disk instability) and was ejected to a large orbit through the gravitational interaction
with another massive body.  Monte Carlo simulations of scattering events show that, given unequal planet
mass ratios, the lower-mass 
planets show a much stronger preference
for outward scattering than the more massive components (\citealt{Veras:2004p19769}; \citealt{Ford:2008p19033}) 
with little dependency on which planet initially had the
wider orbit.  If GSC~06214-00210~b was ejected to its present location
as a result of such an event then it is likely that another object at least as massive ($\sim$14~$M_\mathrm{Jup}$) also 
formed in that system.

Ejections of planets onto wide orbits are not uncommon outcomes in simulations so the discovery of a planetary-mass object
at $\sim$300~AU is perhaps not surprising.  The presence of a 
disk around GSC~06214-00210~b, however, implies that if it underwent
scattering then its disk was not destroyed during point of closest approach with the scatterer.  To our knowledge
there are no studies that specifically investigate the survivability of circumplanetary disks during scattering events.
Here we qualitatively examine the two relevant length scales--- the separation at closest approach during a scattering event 
and the
circumplanetary disk radius--- to assess the likelihood of a disk surviving such an encounter.  

The criterion for orbital stability between two co-planar planets on circular orbits was found by \citet{Gladman:1993p22293} to be

\begin{equation}
\Delta_\mathrm{cr} \geq 2 \sqrt{3} \ R_\mathrm{H,M},
\end{equation}

\noindent where $\Delta$=$a_2 - a_1$ is the difference 
between the initial semimajor axes of the planets
and $R_\mathrm{H,M}$ is the mutual Hill radius (\citealt{Marchal:1982p22296}):

\begin{equation}{\label{eq:rhill}}
R_\mathrm{H,M} = \left( \frac{m_1 + m_2}{3 M_*} \right)^{1/3} \frac{a_1 + a_2}{2}.
\end{equation}

\noindent This radius defines the region in which the gravitational force 
between two bodies is larger than the force on them due to the star.  
The critical range for stability $\Delta_\mathrm{cr}$ can be 
divided into an outer region encompassing weak interactions, which can result in simple rearrangements of the system
architectures with new quasi-stable orbits (\citealt{Ford:2001p22238}; 
\citealt{Veras:2004p19769}), and an inner region where strong interactions occur, which is characterized by chaotic
events made up of collisions and ejections.  This boundary must be found empirically from Monte Carlo simulations
(see \citealt{Gladman:1993p22293} for the empirical relation for equal mass planets).

To test the critical range for stability $\Delta_\mathrm{cr}$, \citet{Chambers:1996p22295} ran simulations of two interacting planets with 
various separations and confirmed
that their orbits were stable for at least 10$^7$~yr when $\Delta$$>$$2\sqrt{3} \ R_\mathrm{H,M}$, and when $\Delta$$<$$2\sqrt{3} \ R_\mathrm{H,M}$ they 
eventually experienced a close encounter
with a distance at closest approach of $<$1~$R_\mathrm{H,M}$.   Although the planet masses 
in that study were smaller than mass scales for a GSC~06214-00210-like system,
a similar study for $\sim$Jovian mass planets by \citet{Marzari:2002p22294} resulted in 
similar close approach scales of $<$1~$R_\mathrm{H,M}$ and a large frequency of ejections.
Most scattering studies focus on various outcomes
as a function of the initial planet spacings
without discussing the distances of closest approach from the simulations, 
which is the relevant length scale we are interested in
here.  The exception is for collisions, defined
by various authors to be the outcome when $\Delta$ becomes less than about the sum of the planetary radii.  Given the
relatively large frequency of collisions in some simulations (e.g., \citealt{Ford:2001p22238} find that collision rates can reach tens of percent
depending on the planet radii and semimajor axes), the distances of closest approach appear to 
be quite small, which is consistent with the separation of $<$1~$R_\mathrm{H,M}$ found by \citet{Chambers:1996p22295} and 
\citet{Marzari:2002p22294}.

Circumplanetary disks are thought to form around young gas giant planets from accretion of circumstellar disk material
once a planet has opened a gap in the protoplanetary disk (e.g., \citealt{Lubow:1999p22013}; \citealt{Ward:2010p21201}).  
In our Solar System the regular satellites of the giant planets are fossil records of these structures (\citealt{Canup:2002p22009}) and
the excess optical emission from Fomalhaut~b has been attributed to scattering from an enormous disk around that extrasolar planet 
(\citealt{Kalas:2008p18842}).
The outer radii of these disks have been investigated by several authors using analytical
arguments and 
hydrodynamical simulations.  
\citet{Quillen:1998p22297}, \citet{Ayliffe:2009p22292}, and \citet{Martin:2011p22270} 
obtain similar results of $\sim$0.3--0.4~$R_\mathrm{H}$, which roughly corresponds the centrifugal
radius.  (Note that the \emph{individual} Hill radius $R_H$ is the limiting case of the mutual Hill radius $R_{H,M}$
when $m_2$ goes to zero and $a_1=a_2$. 
If $m_2 = c \ m_1$ and $a_1 \sim a_2$ then 
$R_\mathrm{H,M}$ and $R_\mathrm{H}$
differ by a factor of $(1+c)^{1/3}$.)
 
Since Hill radii depend on the planet masses and semimajor axes, we consider the 1:1, 1:2, and 1:5 cases
of planet mass ratios to better understand the relative magnitudes of the scales of disk disruption and scattering closest approach.
For an equal-mass coplanar scattering event ($m_2$=$m_1$), the maximum separation at which the disks will not be
disrupted is ($r_\mathrm{disk \ 1}$+$r_\mathrm{disk \ 2}$)$\sim$0.7~$R_\mathrm{H}$.  
Assuming $a_1$$\sim$$a_2$ during a close encounter, the conversion from mutual Hill radius to the point-particle
Hill radius is a factor of 1.26, so the approximate scale of closest approach is $\lesssim$1.26~$R_\mathrm{H}$.  
For the case where $m_2$=$2m_1$, the disk disruption scale is $\sim$0.8~$R_\mathrm{H}$ and 
the distance of closest approach is $\lesssim$1.4~$R_\mathrm{H}$.  Likewise, when $m_2$=$5m_1$,
the disruption scale becomes $\sim$0.95~$R_\mathrm{H}$ and 
the distance of closest approach is $\lesssim$1.8~$R_\mathrm{H}$.

These results suggest that the disruption of circumplanetary disks may be common in scattering events,
especially if the distance to closest approach is less than about one half of a mutual Hill radius.  
The degree to which
a disk is affected by these interactions (ie, disk truncation vs. complete destruction) 
likely depends on the disk sizes, planet mass ratio, relative disk inclinations, relative planet speeds, and impact parameter.
Nevertheless, we interpret the retention of the disk around GSC~06214-00210~b as evidence against
a past scattering event, although detailed simulations are needed.

Further insight into the formation of wide, low-mass companions can be gleaned
from the architectures of other low-mass ratio systems (see \S \ref{sec:genaccretion}).
In a two-planet scattering event involving an ejection, energy conservation requires the remaining bound planet
to have a final semimajor axis greater than one-half of its initial semimajor axis (e.g., \citealt{Ford:2008p19033}).
Since none of the wide planetary-mass companions discovered so far have an
observed scatterer present in the system, the scattering hypothesis appears unlikely,
although the dearth of published detection limits for additional companions in these systems
hampers a more quantitative assessment.  The notable exception is deep AO imaging 
of the 1RXS~J1609--2105 system by \citet{Lafreniere:2010p20986}, and somewhat shallower AO imaging
of the GSC~06214-00210 system by \citet{Ireland:2011p21592}.

\begin{figure}
  \resizebox{\textwidth}{!}{\includegraphics{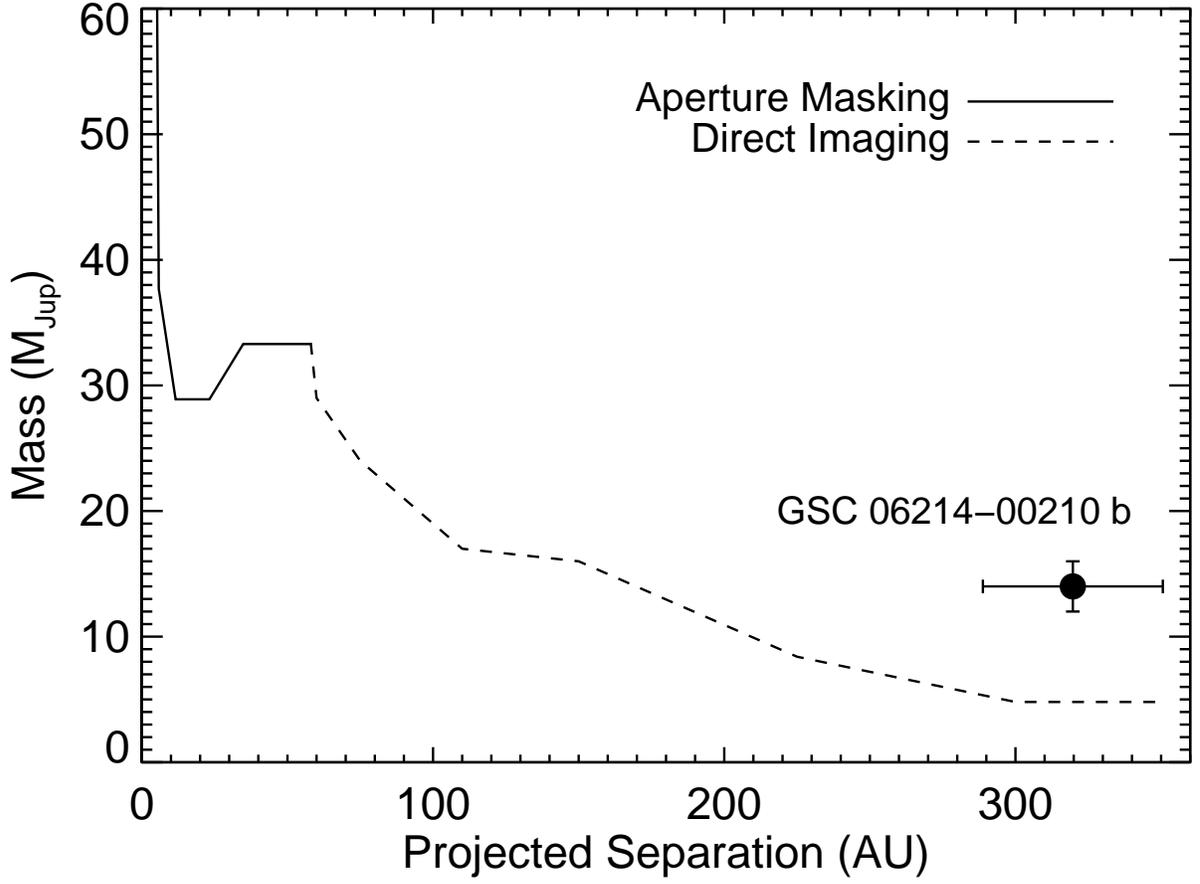}}
  \caption{Constraints on additional companions around GSC~06214-00210 based on aperture masking results from 
  \citet{Kraus:2008p18014} and direct imaging from \citet{Ireland:2011p21592}.  Objects with masses twice that of
  GSC~06214-00210~b are excluded at projected separations $\gtrsim$60~AU.  The aperture masking limits exclude brown dwarfs
  with masses $\gtrsim$35~$M_\mathrm{Jup}$ at projected separations $\gtrsim$5~AU. The uncertainty in the projected
  separation of GSC~06214-00210~b (30~AU) is dominated by the distance error. \label{detlimits} } 
\end{figure}

Finally, we note that if GSC~06214-00210~b was scattered to a large orbit then we can constrain the 
final location of the hypothetical scatterer 
using the aperture masking detection limits from \citet{Kraus:2008p18014} combined with the direct imaging
detection limits from \citet{Ireland:2011p21592}.  As shown in Figure~\ref{detlimits}, the detection limits
exclude objects with the same mass as GSC~06214-00210~b at projected separations $\gtrsim$170~AU,
30~$M_\mathrm{Jup}$ objects (roughly twice the mass of GSC~06214-00210~b) $\gtrsim$60~AU,
and 40~$M_\mathrm{Jup}$ objects $\gtrsim$5~AU.

\subsection{Accretion in Bound and Free-Floating Planetary-Mass Objects}{\label{sec:genaccretion}}

The discovery of strong accretion in GSC~06214-00210~b prompts the broader question of how common
this phenomenon is in other objects.  
Although the census of young planetary-mass companions has been steadily growing over the
past several years, the moderate-resolution spectra needed to detect infrared emission lines 
have only been obtained for a fraction of these objects.

Six companions with masses $\lesssim$20~$M_\mathrm{Jup}$ and ages $\lesssim$10~Myr are known: 
SR~12~C (\citealt{Kuzuhara:2011p21922}), 
CHXR~73~B (\citealt{Luhman:2006p19659}), 1RXS~J1609--2105~b (\citealt{Lafreniere:2008p14057}), 
DH~Tau~b (\citealt{Itoh:2005p19673}), CT~Cha~B (\citealt{Schmidt:2008p14108}), and GSC~06214-00210~b 
(note that the uncertainties in masses and ages can be rather large).
We have ignored companions to young brown dwarfs (e.g., 2M1207~Ab) since the formation of those systems was probably
different from that of stars and wide companions. 
Low-resolution spectra have been published for most of these companions, but moderate-resolution spectra 
have only been acquired for 1RXS~J1609--2105~b (\citealt{Lafreniere:2008p14057}; \citealt{Lafreniere:2010p20986}), 
CT~Cha~B (\citealt{Schmidt:2008p14108}), and GSC~06214-00210~b (this work).
CT~Cha~B and GSC~06214-00210~b both show Pa$\beta$ emission, but there is no evidence of accretion or thermal 
disk emission out to 4~$\mu$m for 1RXS~J1609--2105~b.

Two out of three very low-mass companions therefore show evidence of a disk.  The sample size is minute, but this could 
hint that a large fraction of these objects are accreting.  Moderate resolution near-infrared spectroscopy 
of the remaining companions would help address this matter.  Intriguingly, the low-mass ($\sim$25~$M_\mathrm{Jup}$) 
brown dwarf companion GQ~Lup~B  
also shows Pa$\beta$ emission (\citealt{Seifahrt:2007p21667}; \citealt{Schmidt:2008p14108}; \citealt{Lavigne:2009p19386}),
which may support this notion.

With a large enough sample size, the presence of accretion or thermal disk emission from young planetary-mass 
companions might 
be used to learn about their formation mechanisms.  If this population formed like free-floating planetary-mass objects then
they should both share similar physical properties such as disk frequencies.  
Such a result would be a sign of a common formation mechanism.  
 In the future, a similar analysis might be applied to extrasolar planets
found at close separations (e.g., $<$100~AU) and those at wide separations ($>$100~AU) to test
formation mechanisms, perhaps in the context of core accretion and disk instability.

 \section{Conclusion}

Our $J$- and $H$-band spectroscopy of GSC~06214-00210~b reveals it is 
a late-type object (L0$\pm$1) with several signatures of youth, including very strong Pa$\beta$ emission.
The discovery of accretion in this object confirms
the suggestion by \citet{Ireland:2011p21592} based on its $K-L'$ color that it possesses a circumplanetary disk.
The accretion rate of 10$^{-10.7}$ $M_{\odot}$ yr$^{-1}$ is higher than other objects with comparable masses but is consistent with
the scatter in accretion rates at larger masses.
Atmospheric model fits to our spectrum yield relatively warm temperatures (2700--3000~K), which disagrees with
the cooler predictions from evolutionary models (2200$\pm$100~K).  
With our new luminosity measurement of --3.1$\pm$0.1~dex, we refine the 
predicted mass of
GSC~06214-00210~b to 14$\pm$2~$M_\mathrm{Jup}$, making it the lowest-mass companion 
to harbor a disk.  In addition, our optical and near-infrared spectroscopy of the primary indicate an earlier spectral
type of K7$\pm$0.5 than previously reported.  We revise the mass estimate of GSC~06214-00210 to 0.9$\pm$0.1~$M_{\odot}$
based on our updated temperature and luminosity.  $WISE$ photometry of the primary reveals a marginal (2-$\sigma$) excess
at 22~$\mu$m.

GSC~06214-00210~b is one of only a handful of known companions orbiting stars at several hundred AU with masses 
straddling the brown dwarf/planetary-mass limit.  While we cannot
unambiguously distinguish the formation mechanism of GSC~06214-00210~b,
we suggest that planet-planet scattering is an unlikely explanation.
The small distance at
closest approach between giant planets in a scattering event intimates that circumplanetary disks are probably disrupted during such encounters.  
If this conclusion is bolstered by more detailed dynamical simulations then 
another explanation---  perhaps \emph{in situ} formation---  must be relevant for the population of 
planetary-mass companions on wide orbits.
This also implies that the young free-floating planetary-mass objects with disks are probably
not scattered planets and instead represent the low-mass tail of brown dwarf formation.
 
 \acknowledgments

We thank our anonymous referee for thorough and helpful comments;
Mark Marley, Didier Saumon, and France Allard for making their models available;
Nicolas Lodieu for providing us with the spectra of USco objects; Eric Ford, Matthew Payne, and 
Aaron Boley for productive discussions about planet scattering; Eric Mamajek for the cross-matched $Hipparcos$ catalog; 
Jim Lyke for his support with OSIRIS; and Niall Deacon and Will Best for their assistance during the observations.  
BPB is grateful to Michael Cushing, John Johnson, and Nader Haghighipour for 
helpful discussions. 
BPB and MCL have been supported by NASA grant NNX11AC31G and NSF grant AST09-09222.
ALK has been supported by NASA
through Hubble Fellowship grant 51257.01 awarded by STScI, which is
operated by AURA, INc., for NASA under contract NAS 5-26555.
This research has made use of the NASA/ IPAC Infrared Science Archive, which is operated by the Jet Propulsion Laboratory, California Institute of Technology, under contract with the National Aeronautics and Space Administration.
This publication makes use of data products from the Wide-field Infrared Survey Explorer, which is a joint project of the University of California, Los Angeles, and the Jet Propulsion Laboratory/California Institute of Technology, funded by the National Aeronautics and Space Administration.
We utilized data products from the Two Micron All Sky Survey, which is a joint project of the University of Massachusetts and the Infrared Processing and Analysis Center/California Institute of Technology, funded by the National Aeronautics and Space Administration and the National Science Foundation.
 NASA's Astrophysics Data System Bibliographic Services together with the VizieR catalogue access tool and SIMBAD database 
operated at CDS, Strasbourg, France, were invaluable resources for this work.
The DENIS project has been partly funded by the SCIENCE and the HCM plans of
the European Commission under grants CT920791 and CT940627.
It is supported by INSU, MEN and CNRS in France, by the State of Baden-W\"urttemberg 
in Germany, by DGICYT in Spain, by CNR in Italy, by FFwFBWF in Austria, by FAPESP in Brazil,
by OTKA grants F-4239 and F-013990 in Hungary, and by the ESO C\&EE grant A-04-046.
Finally, mahalo nui loa to the kama`\={a}ina of Hawai`i for their support of Keck and the Mauna Kea observatories.
We are grateful to conduct observations from this mountain.

\facility{{\it Facilities}: Keck:II (OSIRIS), IRTF (SpeX), UH:2.2m (SNIFS) }

\newpage


\clearpage

\newpage

\begin{deluxetable}{lc}
\tablewidth{0pt}
\tablecolumns{2}
\tablecaption{Properties of the Primary GSC 06214-00210\label{tab:priprop}}
\tablehead{
        \colhead{Property}   &    \colhead{Value}    
        }   
\startdata
\cutinhead{Observed}
$SpT$     &     K7$\pm$0.5      \\
$L_\mathrm{Bol}$ & 0.38 $\pm$ 0.07 $L_{\odot}$ \\
$r'$  (CMC14) & 11.94 mag \\
$i$ (DENIS) & 11.08 $\pm$ 0.03 mag \\
$J$ (DENIS) & 10.05 $\pm$ 0.06 mag \\
$K_S$ (DENIS) & 9.31 $\pm$ 0.06 mag \\
$J$ (2MASS) & 9.998 $\pm$ 0.027 mag \\
$H$ (2MASS) & 9.342 $\pm$ 0.024 mag \\
$K_S$ (2MASS) & 9.152 $\pm$ 0.021 mag \\
3.4~$\mu$m (WISE) &  9.094 $\pm$ 0.024 mag \\
4.6~$\mu$m (WISE) &  9.111 $\pm$ 0.021 mag \\
12~$\mu$m (WISE) &  8.985 $\pm$ 0.035 mag \\
22~$\mu$m (WISE) &  8.315 $\pm$ 0.271 mag \\

\cutinhead{Estimated}
Distance  & 145 $\pm$ 14 pc \\
Age & 5 $\pm$2 Myr \\
Mass  &   0.9 $\pm$ 0.1  $M_{\odot}$    \\
$T_\mathrm{eff}$ &  4200 $\pm$ 150 K \\

\enddata
\end{deluxetable}

\begin{deluxetable}{lc}
\tablewidth{0pt}
\tablecolumns{2}
\tablecaption{Properties of the Companion GSC 06214-00210~b\label{tab:compprop}}
\tablehead{
        \colhead{Property}   &    \colhead{Value}    
        }   
\startdata
\cutinhead{Observed}
$SpT$   & L0$\pm$1 \\
Proj. Sep.  & 2$\farcs$2 (320 $\pm$ 30 AU) \\
$EW$(Pa$\beta$) & --11.4 $\pm$ 0.3 \AA \\
log($L_\mathrm{Bol}$/$L_{\odot}$) & --3.1 $\pm$ 0.1 \\
$J$ (MKO)\tablenotemark{a} & 16.25 $\pm$ 0.04 mag \\
$H$ (MKO)\tablenotemark{a} & 15.55 $\pm$ 0.04 mag \\
$K$ (MKO)\tablenotemark{a,b} & 14.94 $\pm$ 0.03 mag \\
$L'$\tablenotemark{a}  & 13.75 $\pm$ 0.07 mag \\

\cutinhead{Estimated}
Mass (Evol.)  &  14 $\pm$ 2 $M_\mathrm{Jup}$ \\
$T_\mathrm{eff}$ (Evol.) &  2200 $\pm$ 100 K \\
$T_\mathrm{eff}$ (BT-Settl) &  2700 $\pm$ 200 K \\
log($L$(Pa$\beta$)/$L_{\odot}$) & --6.14 $\pm$ 0.08 \\
log($L_\mathrm{acc}$/$L_{\odot}$) & --4.4 $\pm$ 1.3 \\
log(\emph{\.{M}}/$M_{\odot}$ yr$^{-1}$ ) & --10.7 $\pm$ 1.3 \\

\enddata
\tablenotetext{a}{Photometry from \citet{Ireland:2011p21592}.}
\tablenotetext{b}{Weighted mean and uncertainty of four measurements from \citet{Ireland:2011p21592}.}
\end{deluxetable}

\clearpage

\end{document}